\documentclass[pdflatex,sn-basic]{sn-jnl}

\usepackage{bm}
\usepackage{amssymb}
\usepackage{amsmath}
\usepackage{enumitem}
\usepackage{CJK}            
\catcode`\|=12\relax        
\usepackage{orcidlink}
\bibpunct{(}{)}{;}{a}{}{,}

\usepackage{etoolbox}

\makeatletter

\patchcmd{\NAT@citex}
  {\@citea\NAT@hyper@{%
     \NAT@nmfmt{\NAT@nm}%
     \hyper@natlinkbreak{\NAT@aysep\NAT@spacechar}{\@citeb\@extra@b@citeb}%
     \NAT@date}}
  {\@citea\NAT@nmfmt{\NAT@nm}%
   \NAT@aysep\NAT@spacechar\NAT@hyper@{\NAT@date}}{}{}

\patchcmd{\NAT@citex}
  {\@citea\NAT@hyper@{%
     \NAT@nmfmt{\NAT@nm}%
     \hyper@natlinkbreak{\NAT@spacechar\NAT@@open\if*#1*\else#1\NAT@spacechar\fi}%
       {\@citeb\@extra@b@citeb}%
     \NAT@date}}
  {\@citea\NAT@nmfmt{\NAT@nm}%
   \NAT@spacechar\NAT@@open\if*#1*\else#1\NAT@spacechar\fi\NAT@hyper@{\NAT@date}}
  {}{}

\makeatother

\newcommand {\eV}           {\,\rm eV}

\newcommand {\kms}          {\,\rm km\,\rm s^{-1}}
\newcommand {\pc}           {\,\rm pc}
\newcommand {\kpc}          {\,\rm kpc}
\newcommand {\Mpc}          {\,\rm Mpc}

\newcommand {\kpch}         {\,h^{-1}\rm kpc}
\newcommand {\Mpch}         {\,h^{-1}\rm Mpc}
\newcommand {\Myr}          {\,\rm Myr}
\newcommand {\Gyr}          {\,\rm Gyr}
\newcommand {\mFDM}         {m_{22}}
\newcommand {\Msun}         {\,\rm{M}_{\odot}}
\newcommand {\Msunh}        {\,h^{-1}\rm{M}_{\odot}}
\newcommand {\Mh}           {M_{\rm h}}
\newcommand {\rh}           {r_{\rm h}}
\newcommand {\kJ}           {k_{\rm J}}
\newcommand {\kJphy}        {k_{\rm J, phy}}
\newcommand {\kJeq}         {k_{\rm J, eq}}
\newcommand {\khf}          {k_{1/2}}
\newcommand {\Mhf}          {M_{1/2}}
\newcommand {\rhob}         {\tilde{\rho}_{\rm b}}
\newcommand {\rhos}         {\rho_{\rm s}}
\newcommand {\rs}           {r_{\rm s}}
\newcommand {\Ms}           {M_{\rm s}}
\newcommand {\Es}           {E_{\rm s}}
\newcommand {\LdB}          {\lambda_{\rm dB}}
\newcommand {\vdisp}        {\sigma_{\rm 1D}}
\newcommand {\Dh}           {\Delta h}
\newcommand {\Dt}           {\Delta t}
\newcommand {\dR}           {\delta R}
\newcommand {\dI}           {\delta I}
\newcommand {\dRcom}        {\delta \tilde{R}}
\newcommand {\dIcom}        {\delta \tilde{I}}
\newcommand {\ave}[1]       {\left\langle {#1} \right\rangle}

\newcommand*{\textcode}[1]  {\textsc{\small #1}}

\newcommand*{\sref}[1]      {Section~\ref{#1}}
\newcommand {\fref}[1]      {Fig.~\ref{#1}}

\newcommand {\eref}[1]      {Eq.~(\ref{#1})}
\newcommand {\erefp}[1]     {Eq.~\ref{#1}}

\begin{document}

\title[FDM simulations]{Fuzzy dark matter simulations}


\author*[1,2,3,4]{\fnm{Hsi-Yu} \sur{Schive}\orcidlink{0000-0002-1249-279X}}
\email{hyschive@phys.ntu.edu.tw}

\affil[1]{\orgdiv{Institute of Astrophysics},      \orgname{National Taiwan University},               \orgaddress{\city{Taipei}, \postcode{10617}, \country{Taiwan}}}
\affil[2]{\orgdiv{Department of Physics},          \orgname{National Taiwan University},               \orgaddress{\city{Taipei}, \postcode{10617}, \country{Taiwan}}}
\affil[3]{\orgdiv{Center for Theoretical Physics}, \orgname{National Taiwan University},               \orgaddress{\city{Taipei}, \postcode{10617}, \country{Taiwan}}}
\affil[4]{\orgdiv{Physics Division},               \orgname{National Center for Theoretical Sciences}, \orgaddress{\city{Taipei}, \postcode{10617}, \country{Taiwan}}}

\abstract{
Fuzzy dark matter (FDM), composed of ultralight bosons, exhibits intricate wave phenomena on galactic scales. Compared to cold dark matter, FDM simulations are significantly more computationally demanding due to the need to resolve the de Broglie wavelength and its rapid oscillations. In this review, we first outline the governing equations and distinctive features of FDM. We then present a range of numerical algorithms for both wave- and fluid-based simulations, discuss their respective advantages and limitations, and highlight representative test problems. To facilitate code comparison, we also provide publicly available initial condition files for both isolated-halo and cosmological simulations.
}


\maketitle

\pagestyle{myheadings}
\markright{H.-Y. Schive}
{
\small
\setcounter{tocdepth}{3}
\tableofcontents
}

\section{Introduction}
\label{sec:intro}

Fuzzy dark matter (FDM) offers a compelling alternative to cold dark matter (CDM). It consists of ultralight bosons characterized by a single parameter---the boson mass $\mFDM \equiv m/10^{-22}\eV \sim 1\text{--}10^3$. The associated large de Broglie wavelength gives rise to distinctive wave-like phenomena on galactic scales, including the suppression of low-mass halos due to quantum pressure, density fluctuations and vortices from wave interference, and the formation of compact soliton cores as the ground state of the host halo potential. For comprehensive reviews on the theoretical background and observational constraints of FDM, see, for example, \citet{Marsh2016, Niemeyer2020, Hui2021b, Chavanis2025, Eberhardt2025}.

This review concentrates on computational aspects of FDM research \citep[for an earlier review, see][]{Zhang2019}. Numerical simulations play an indispensable role in identifying unique FDM properties and confronting model predictions with data. For example, FDM simulations have been used to study the nonlinear evolution of large-scale structure, the halo mass function, the dynamical heating of stellar systems caused by wave interference, and the formation and dynamics of soliton cores. Here we focus on the simplest scenario: a single FDM species without self-interaction, although the simulation techniques discussed can be readily generalized to explore such extensions.

The remainder of this review is organized as follows. First, we introduce the governing equations and general features of FDM in \sref{sec:intro}. We then present a variety of FDM algorithms in \sref{sec:methods} and describe the associated numerical challenges in \sref{sec:challenges}. Representative numerical tests and results are provided in \sref{sec:test}. Finally, we conclude in \sref{sec:conclusions}. Links to the initial condition files for the isolated-halo and cosmological simulations are made available in \nameref{sec:data}.

\subsection{Governing equations}
\label{subsec:intro_eq}

We present the wave formulation of FDM in \sref{subsec:wave_eq} and its fluid representation in \sref{subsec:fluid_eq}.

\subsubsection{Wave formulation}
\label{subsec:wave_eq}

The governing equations of FDM are the Schr\"odinger--Poisson equations:
\begin{align}
  i\frac{\partial \psi}{\partial t} &= \left(-\frac{\hbar}{2m}\nabla^2 + \frac{m}{\hbar} V\right) \psi,
  \label{eq:schroedinger} \\
  \nabla^2 V &= 4\pi G \|\psi\|^2,
  \label{eq:poisson}
\end{align}
where $\psi$ is the wave function, $\hbar$ is the reduced Planck constant, $V$ is the gravitational potential, and $G$ is the gravitational constant. This system of equations is invariant under the following scale transformation \citep{Ruffini1969, Guzman2004}:
\begin{equation}
  (m, \bm{r}, t, \psi, V) \to (\beta m, \alpha^{-1}\beta^{-1/2}\bm{r}, \alpha^{-2}t, \alpha^2\psi, \alpha^2\beta^{-1}V),
  \label{eq:scale_transf}
\end{equation}
where $\alpha$ and $\beta$ are arbitrary dimensionless constants. Note also that $m$ and $\hbar$ always appear as the ratio $m/\hbar$.

For cosmological simulations, one can define $\bm{\tilde{r}}=a^{-1}\bm{r}$, $\tilde{\psi}=a^{3/2}e^{-imHr^2/2\hbar}\psi$, and $\tilde{V}=a^2(V-V_b)$ to rewrite Eqs. (\ref{eq:schroedinger}--\ref{eq:poisson}) in comoving coordinates:
\begin{align}
  ia^2\frac{\partial \tilde{\psi}}{\partial t} &= \left(-\frac{\hbar}{2m}\tilde{\nabla}^2 + \frac{m}{\hbar} \tilde{V}\right) \tilde{\psi},
  \label{eq:schroedinger_com} \\
  \tilde{\nabla}^2 \tilde{V} &= 4\pi G a\left( \|\tilde{\psi}\|^2 - \rhob \right),
  \label{eq:poisson_com}
\end{align}
where $a$ is the scale factor, $H$ is the Hubble parameter, $V_b$ is the potential associated with the homogeneous background, and $\rhob$ is the comoving background density. The similarities between Eqs. (\ref{eq:schroedinger}--\ref{eq:poisson}) and (\ref{eq:schroedinger_com}--\ref{eq:poisson_com}) allows the same numerical algorithms to be applied to both cases, particularly if one further introduces the so-called `supercomoving time', $d\tilde{t}=a^{-2}dt$ \citep{Martel1998}.

The wave function $\psi$ can be expressed in polar coordinates:
\begin{equation}
  \psi = R+iI = \rho^{1/2} e^{i S},
\label{eq:wave_polar}
\end{equation}
where $R=\Re({\psi})$ and $I=\Im(\psi)$ are the real and imaginary parts of the wave function, respectively, and $S$ is the phase. One can then derive physical quantities, such as mass density $\rho$, bulk velocity $\bm{v}$, `thermal' velocity $\bm{w}$, and energy density $e$, as
\begin{align}
  \rho &= \|\psi\|^2,
  \label{eq:mass_dens} \\
  \bm{v} &= \frac{\hbar}{m} \frac{R\bm{\nabla}I - I\bm{\nabla}R}{\|\psi\|^2} = \frac{\hbar}{m} \bm{\nabla} S,
  \label{eq:bulk_vel} \\
  \bm{w} &= \frac{\hbar}{m} \frac{R\bm{\nabla}R + I\bm{\nabla}I}{\|\psi\|^2} = \frac{\hbar}{2m} \frac{\bm{\nabla}\rho}{\rho},
  \label{eq:thermal_vel} \\
  e &= \frac{\hbar^2}{2m^2} \| \bm{\nabla}\psi \|^2 = \frac{1}{2}\rho(v^2+w^2).
  \label{eq:engy_dens}
\end{align}
Similarly, in comoving coordinates, we can define $\tilde{\psi} = \tilde{\rho}^{1/2} e^{i \tilde{S}}$, with $\tilde{\rho}=a^3\rho$ being the comoving mass density. The peculiar velocity is then given by
\begin{equation}
  \bm{v_{\rm pec}} = a^{-1} \frac{\hbar}{m} \bm{\tilde{\nabla}} \tilde{S}.
  \label{eq:peculiar_vel}
\end{equation}

\subsubsection{Fluid formulation}
\label{subsec:fluid_eq}

From the transformation in \eref{eq:wave_polar}, \eref{eq:schroedinger} can be recast into the Hamilton--Jacobi--Madelung equations:
\begin{align}
  \frac{m}{\hbar}\frac{\partial \rho}{\partial t} + \bm{\nabla} \cdot (\rho \bm{\nabla} S) &= 0,
  \label{eq:hjm_continuity} \\
  \frac{m}{\hbar}\frac{\partial S}{\partial t} + \frac{1}{2}\|\bm{\nabla} S\|^2
  + \frac{m^2}{\hbar^2} \left( Q + V \right) &= 0,
  \label{eq:hjm_phase}
\end{align}
where $Q$ is referred to as the `quantum potential':
\begin{equation}
  Q = - \frac{\hbar^2}{2m^2} \frac{{\nabla}^2 \sqrt{\rho}}{\sqrt{\rho}}
    = - \frac{\hbar^2}{2m^2}\left[ \frac{\nabla^2\rho}{2\rho} - \frac{\|\bm{\nabla}\rho\|^2}{4\rho^2} \right].
  \label{eq:quantum_potential}
\end{equation}
By taking the gradient of \eref{eq:hjm_phase} and defining the bulk velocity $\bm{v}$ from \eref{eq:bulk_vel}, one obtains the Madelung equations:
\begin{align}
  \frac{\partial \rho}{\partial t} + \bm{\nabla} \cdot \left(\rho \bm{v}\right) &= 0,
  \label{eq:madelung_continuity} \\
  \frac{\partial \bm{v}}{\partial t} + \bm{v} \cdot \bm{\nabla} \bm{v}
  + \bm{\nabla} (Q+V) &= 0.
  \label{eq:madelung_velocity}
\end{align}
Note that $\bm{v} \propto \bm{\nabla}S$ is a gradient flow, so the flow vorticity $\bm{\nabla} \times \bm{v}$ vanishes everywhere except at the location of vortices, where the density drops to zero and both the phase and velocity become ill-defined (see \sref{subsec:challenges_vortices}).

\eref{eq:madelung_velocity} can be rewritten into a conservative form:
\begin{equation}
  \frac{\partial \rho\bm{v}}{\partial t} + \bm{\nabla} \cdot \left(\rho\bm{v}\otimes\bm{v} + \bm{\Sigma}\right)
  + \rho\bm{\nabla}V = 0,
  \label{eq:madelung_momentum}
\end{equation}
where $\bm{\Sigma}$ is the `quantum stress tensor', often also referred to as the `quantum pressure':
\begin{equation}
  \Sigma_{ij}
    = \frac{\hbar^2}{4m^2}\left( \frac{1}{\rho} \frac{\partial\rho}{\partial x_i}\frac{\partial\rho}{\partial x_j}
      - \delta_{ij}\nabla^2\rho \right)
      \quad\text{or}\quad
      - \frac{\hbar^2}{4m^2}\left( \rho \frac{\partial^2 \ln\rho}{\partial x_i \partial x_j} \right),
  \label{eq:quantum_pressure}
\end{equation}
with $\delta_{ij}$ denoting the Kronecker delta. The two forms of $\Sigma_{ij}$ are equivalent after taking the divergence. \eref{eq:madelung_momentum} is analogous to the Euler equations of hydrodynamics, but with a key distinction: the quantum pressure term introduces several unique features in FDM (see \sref{subsec:fdm_features}).

Similar to the comoving Schr\"odinger equation \eref{eq:schroedinger_com}, Eqs. (\ref{eq:hjm_continuity}--\ref{eq:quantum_pressure}) can be converted to comoving coordinates by replacing $(\rho, S, \bm{v}, V, \bm{\nabla}, dt)$ with $(\tilde{\rho}, \tilde{S}, \bm{\tilde{v}}, \tilde{V}, \bm{\tilde{\nabla}}, d\tilde{t})$, where $\bm{\tilde{v}} = (\hbar/m)\bm{\tilde{\nabla}}\tilde{S} = a\bm{v_{\rm pec}}$.

\subsection{FDM features}
\label{subsec:fdm_features}

This subsection highlights key FDM features: the suppression of small-scale structure (\sref{subsec:small_scale_suppression}), soliton cores (\sref{subsec:soliton}), and density granulation (\sref{subsec:granule}).

\subsubsection{Suppression of small-scale structure}
\label{subsec:small_scale_suppression}

Quantum pressure in FDM counteracts gravity and introduces a Jeans scale, which is associated with the de Broglie wavelength of the ground state in a potential well. Structures larger than this scale grow similarly to CDM, whereas smaller structures are suppressed due to enhanced velocity arising from the uncertainty principle. The comoving Jeans wavenumber $\kJ$ remains constant during the radiation-dominated era and increases slowly as $\kJ \propto \mFDM^{1/2}a^{1/4}$ during the matter-dominated era \citep{Hu2000}. This slowly varying $\kJ$ leads to a sharp transition in the linear matter power spectrum at $k \sim \kJeq$, where $\kJeq = 9\,\mFDM^{1/2}\Mpc^{-1}$ is the Jeans wavenumber at matter-radiation equality.

The suppression of the FDM power spectrum $P_{\rm FDM}(k,z)$ relative to CDM can be expressed as
\begin{equation}
  P_{\rm FDM}(k,z) = T_{\rm FDM}^2(k,z)P_{\rm CDM}(k,z),
  \label{eq:transfer_func1}
\end{equation}
where $T_{\rm FDM}$ is the FDM transfer function and $z$ is the redshift. \citet{Hu2000} proposed the following analytical fitting formula for $T_{\rm FDM}$ in the matter-dominated era:
\begin{equation}
  T_{\rm FDM} \approx \frac{\cos x^3}{1+x^8},\;\; x=1.61\,\mFDM^{1/18}\frac{k}{\kJeq}.
  \label{eq:transf_func2}
\end{equation}
It is modeled as redshift-independent because the strong suppression at $k \sim \kJeq$ is primarily determined during the radiation-dominated era. Alternatively, $T_{\rm FDM}$ can be computed directly using the Boltzmann code \textcode{axionCAMB} \citep{Hlozek2015}.

From \eref{eq:transf_func2}, one can define the `half-mode' wavenumber $\khf$, where $T_{\rm FDM}(\khf) = 1/2$, and the associated halo mass $\Mhf = 4\pi(\pi/\khf)^3\rhob/3$ as
\begin{equation}
  \khf \approx 5.1\,\mFDM^{4/9}\Mpc^{-1},\;\;
  \Mhf \approx 3.8\times10^{10}\,\mFDM^{-4/3} \Msun.
  \label{eq:half_mode}
\end{equation}
$\Mhf$ characterizes the mass scale below which the FDM halo mass function, $\left.{\rm d}n/{\rm d}\Mh\right\rvert_{\rm FDM}$, is significantly suppressed compared to CDM. \citet{Schive2016} provided the following fitting function:
\begin{equation}
  \left.\frac{{\rm d}n}{{\rm d}\Mh}\right\rvert_{\rm FDM}(\Mh,z) =
  \left.\frac{{\rm d}n}{{\rm d}\Mh}\right\rvert_{{\rm CDM}}(\Mh,z)
  \left[ 1 + \left( \frac{\Mh}{0.42\Mhf} \right)^{-1.1} \right]^{-2.2}.
  \label{eq:mass_function}
\end{equation}
Similar to \eref{eq:transf_func2}, the suppression term $[ 1 + (\Mh/0.42\Mhf)^{-1.1}]^{-2.2}$ is redshift-independent, as $\kJ > \kJeq$ throughout the matter-dominated era. Furthermore, although \eref{eq:mass_function} was originally derived from collisionless $N$-body simulations with FDM initial conditions (see \sref{subsec:methods_nbody}), it has been confirmed to agree well with genuine FDM wave simulations solving Eqs. (\ref{eq:schroedinger_com}--\ref{eq:poisson_com}) \citep{May2022}. This agreement demonstrates that the FDM halo mass function in cosmological simulations is governed primarily by the initial conditions, with dynamical effects of quantum pressure playing a subdominant role.

\begin{figure*}[th!]
\includegraphics[width=\textwidth]{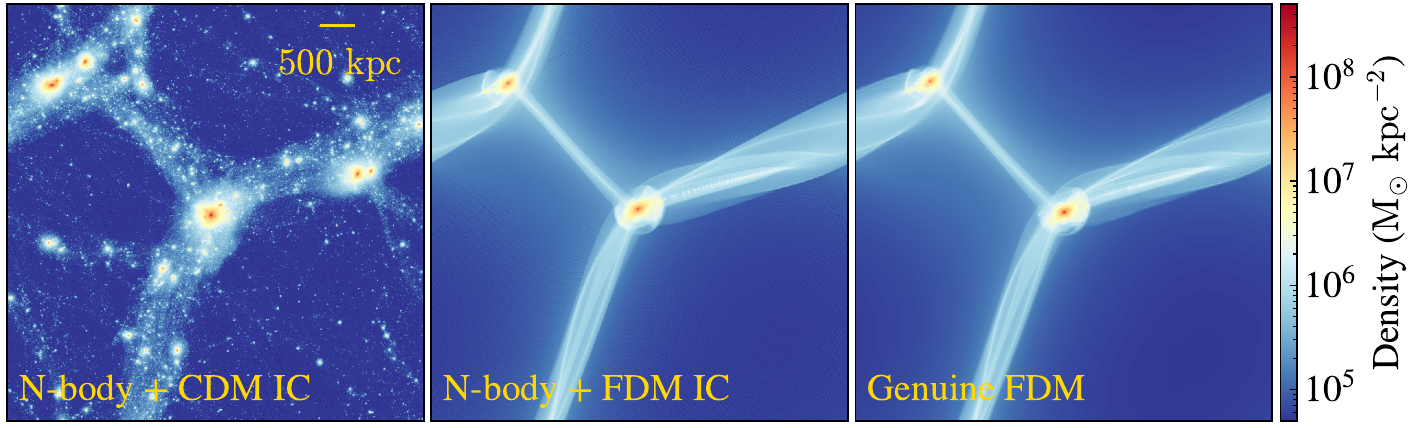}
\caption{
Projected density distributions from three cosmological simulations at $z=0$: (left) a genuine CDM simulation, (middle) a collisionless $N$-body simulation from an FDM initial condition (IC) with $\mFDM=0.1$, and (right) a genuine FDM simulation solving the Schr\"odinger--Poisson equations and their fluid formulation using the same FDM initial condition as the middle panel. Low-mass halos in the latter two cases are significantly suppressed in a very consistent manner, suggesting that the initial conditions dominate over the dynamical effects of quantum pressure in determining the FDM halo mass function
}
\label{fig:cdm_vs_fdm}
\end{figure*}

\fref{fig:cdm_vs_fdm} illustrates these features by comparing the density distributions at $z=0$ from three distinct simulations: collisionless $N$-body simulations initialized with CDM and FDM ($\mFDM = 0.1$) power spectra, and a genuine FDM simulation using a hybrid fluid--wave algorithm (see \sref{subsec:methods_hybrid}) with the same FDM initial condition. The simulation box size is $5.9\Mpc$. The genuine CDM simulation exhibits significantly more low-mass halos. By contrast, the halo distributions in the latter two simulations are similar, although only the genuine FDM simulation can resolve small-scale interference fringes and solitons (see \sref{subsec:soliton} and \sref{subsec:granule}). The two surviving massive halos have a mass of $\Mh=4.5\times10^{11}\Msun$ and $9.1\times10^{11}\Msun$, respectively, close to the half-mode mass $\Mhf \approx 8.2\times10^{11}\Msun$. The $N$-body and FDM simulations are performed using the codes \textcode{GADGET-2} \citep{Springel2005} and \textcode{GAMER} \citep{Schive2018a}, respectively.

The suppression of small-scale structure provides a powerful means to constrain $\mFDM$. Examples include the Lyman-alpha forest power spectrum \citep{Irsic2017, Armengaud2017, Kobayashi2017, Leong2019, Nori2019, Rogers2021}, the cosmic microwave background \citep{Hlozek2015, Hlozek2018}, the subhalo mass function \citep{Du2017b, Benito2020, Schutz2020, Banik2021, Nadler2021, Nadler2024}, the halo abundance and the faint end of the luminosity function at high redshifts \citep{Bozek2015, Schive2016, Corasaniti2017, Menci2017, Schive2018b, Ni2019, Winch2024, Sipple2025}, the galaxy shear weak lensing power spectrum \citep{Dentler2021}, and the 21-cm signal during cosmic dawn \citep{Hotinli2022}.

\subsubsection{Solitons}
\label{subsec:soliton}

\begin{figure*}[th!]
\includegraphics[width=\textwidth]{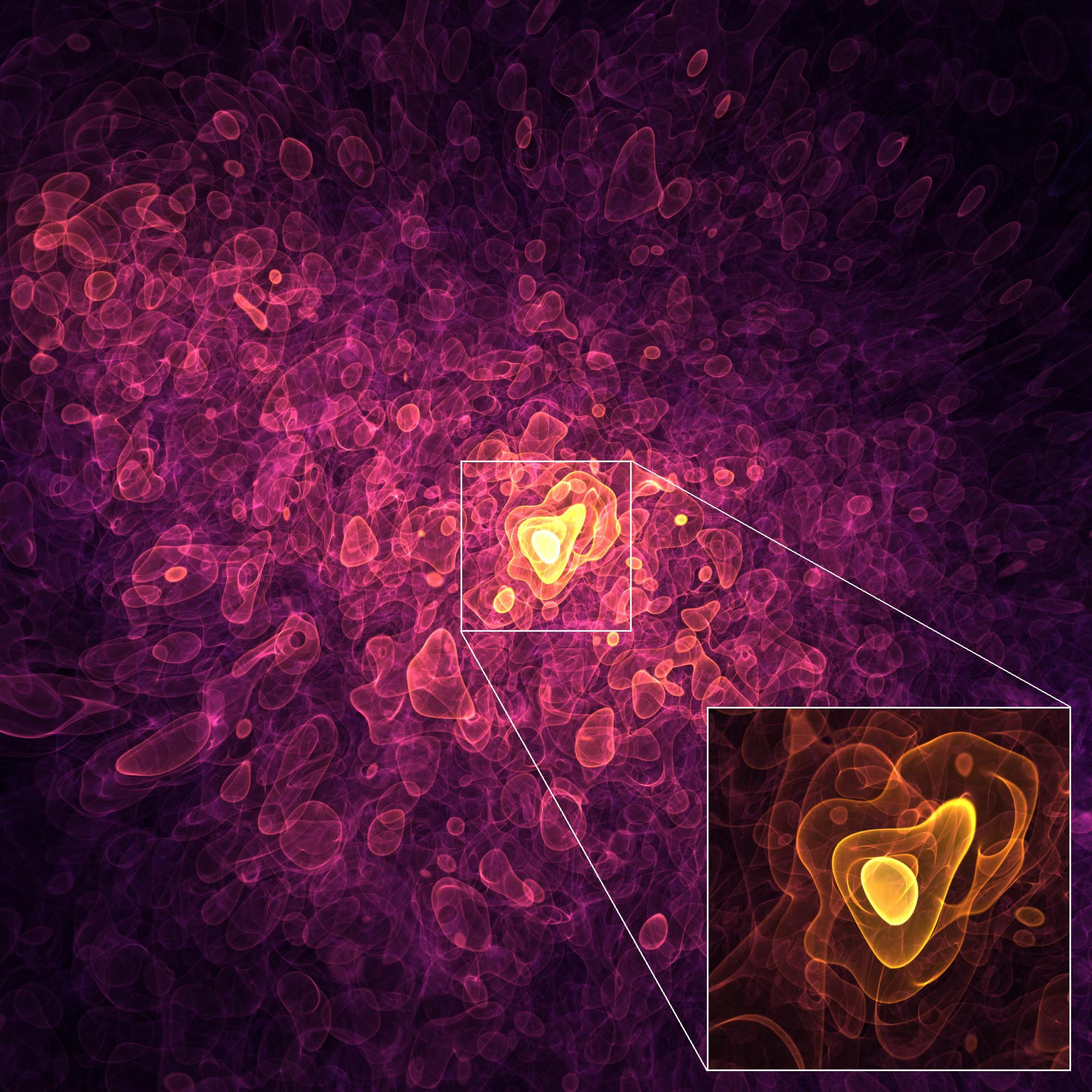}
\caption{
Density distribution of a $5.7\times10^{10}\Msun$ FDM halo with $\mFDM=0.2$ at $z=0$. The halo exhibits ubiquitous, stochastically fluctuating granular structures, with a characteristic granule size determined by the local de Broglie wavelength. The brightest region in the inset highlights the central dense soliton, which has a mass of $5.2\times10^8\Msun$. The soliton radius is comparable to that of the surrounding granules, suggesting thermal equilibrium
}
\label{fig:soliton_granule}
\end{figure*}

FDM halos feature a dense, stable soliton core surrounded by a Navarro-Frenk-White \citep[NFW;][]{Navarro1996} halo permeated by fluctuating density granules \citep{Schive2014a, Schive2014b, Marsh2015a, Mocz2017, Veltmaat2018}. \fref{fig:soliton_granule} displays such a soliton--halo system. The soliton, forming shortly after halo collapse, represents the ground-state solution of the halo potential. The soliton density profile can be well fitted by \citep{Schive2014a, Marsh2015a}
\begin{equation}
  \rhos(r) = \frac{1.95\times10^7\mFDM^{-2}(\rs/\kpc)^{-4}} {[1+9.06\times10^{-2}(r/\rs)^2]^8} \Msun\kpc^{-3},
  \label{eq:soliton_profile}
\end{equation}
which is redshift-independent as the soliton is an object detached from cosmological expansion. See \fref{fig:test_soliton} for an illustration. Here, $\rs$ is the soliton radius defined by $\rhos(\rs)=\rhos(0)/2$. The corresponding soliton mass $\Ms$ within $\rs$ is
\begin{equation}
  \Ms = 5.4\times10^7 \mFDM^{-2} \left(\frac{\rs}{\kpc}\right)^{-1} \Msun.
  \label{eq:soliton_mass_radius}
\end{equation}
Analytical expressions for the enclosed mass and gravitational potential profiles can be found in Eq. (A1) of \citet{Chen2017} and Eq. (6) of \citet{Chiang2021}, respectively. Note that Eqs. (\ref{eq:soliton_profile}--\ref{eq:soliton_mass_radius}) obey the scale transformation
\begin{equation}
  (\rs, \rhos, \Ms, \Es) \to (\alpha^{-1}\rs, \alpha^{4}\rhos, \alpha\Ms, \alpha^3\Es),
  \label{eq:soliton_scale_transf}
\end{equation}
which follows directly from \eref{eq:scale_transf} with $\beta=1$, where $\Es$ denotes the soliton energy. Accordingly, for a fixed FDM particle mass, all soliton solutions form a one-parameter family.

Soliton properties can be inferred from their host halos, known as the soliton--halo relation. Its exact form remains under debate, with proposed scalings including $\Ms \propto \Mh^{1/3}$ \citep{Schive2014b, Du2017a, Veltmaat2018, Chavanis2019} and $\Ms \propto \Mh^{5/9}$ \citep{Mocz2017, Nori2021}. This relation can be interpreted as a manifestation of thermal equilibrium between the soliton and its surrounding halo granules \citep{Liao2025}.  The transition from soliton to halo occurs at $r \approx 3.3\text{--}3.5\,\rs$ \citep{Mocz2017, Chiang2021}, beyond which the average radial density profile follows the NFW model and closely matches the result from a collisionless $N$-body simulation using the same FDM initial condition \citep[][see also the middle and right panels in \fref{fig:cdm_vs_fdm}]{Liao2025}. In contrast to CDM, the concentration parameter of FDM halos decreases for $\Mh \lesssim \Mhf$, as a consequence of the delayed onset of halo formation \citep{Laroche2022, Kawai2024}.

The flat core and mass excess associated with the central soliton is a distinct prediction of FDM, setting it apart from other dark matter scenarios and providing a stringent test of the model. For instance, one can constrain the soliton profile from the size and kinematic data of dwarf galaxies \citep{Schive2014a, Calabrese2016, Chen2017, GonzalezMorales2017, Burkert2020, Safarzadeh2020, Hayashi2021, Zimmermann2025}, ultra-diffuse galaxies \citep{Wasserman2019, Montes2024}, and the central molecular zone of the Milky Way \citep{Li2020}. The deep gravitational potential of a dense compact soliton can (i) accelerate gas accretion and thereby boost the growth of supermassive black holes in the early universe \citep{Chiu2025}, and (ii) produce a characteristic peak in galaxy rotation curves at $r \approx 1.96\,\rs$ with a peak velocity $v_{\rm max} \approx 36\,\mFDM(\Ms/10^8\Msun)\kms$ \citep{Bernal2018, Bar2018, Bar2019, Craciun2020, Chan2021, Bar2022, Khelashvili2023}. This additional potential can also counteract tidal forces, making FDM subhalos more resilient to tidal disruption compared to their CDM counterparts \citep{Chan2025}. On the other hand, if the tidal radius is comparable to $\rs$, solitons become unstable and undergo runaway disruption \citep{Du2018}. Furthermore, solitons exhibit density oscillations and random walk due to wave interference \citep{Veltmaat2018, Schive2020, Li2021, Chiang2021, Chowdhury2021, Zagorac2022, Chiueh2023}, which may dynamically heat or even tidally disrupt nearby stellar systems \citep{Marsh2019, Schive2020, Chowdhury2023, Teodori2026}.

\subsubsection{Density granulation}
\label{subsec:granule}

As shown in \fref{fig:soliton_granule}, another distinct feature of FDM halos is the presence of ubiquitous, stochastically fluctuating granular structures, arising from constructive and destructive interference. The characteristic granule size is set by the local de Broglie wavelength $\LdB$ \citep{Veltmaat2018, BarOr2019, Chowdhury2021},
\begin{equation}
  d_{\rm gra} \approx 0.25\,\LdB \approx 0.3\,\mFDM^{-1}\left( \frac{\vdisp}{100\kms} \right)^{-1} \kpc,
  \label{eq:granule_size}
\end{equation}
where $\LdB=h/m\vdisp$ and $\vdisp$ is the one-dimensional (bulk) velocity dispersion. The characteristic granule lifetime is estimated as
\begin{equation}
  T_{\rm gra} \approx \frac{d_{\rm gra}}{\vdisp} \approx 2.9\,\mFDM^{-1}\left( \frac{\vdisp}{100\kms} \right)^{-2} \Myr.
  \label{eq:granule_time}
\end{equation}
The dynamical effects of these density fluctuations can be modeled by treating FDM granules as quasiparticles \citep{BarOr2019, Church2019, ElZant2020, Chavanis2021, Chowdhury2021}. Assuming a Maxwellian velocity distribution, the effective particle mass is
\begin{equation}
  M_{\rm gra} \approx \frac{\pi\rho d^3}{6},
  \label{eq:granule_mass}
\end{equation}
where $\rho$ denotes the local dark matter density.

The granule size remains approximately constant in the vicinity of the soliton but increases with radius beyond this region, indicating an isothermal velocity distribution in the inner halo and a non-isothermal distribution in the outskirts. This radial variation in temperature correlates positively with the FDM halo concentration parameter \citep{Liao2025}. Furthermore, in contrast to the central soliton, which is supported by quantum pressure with $w \gg v$, the region outside the soliton exhibits $w \sim v$, indicating energy equipartition \citep{Chowdhury2021}. Since $\bm{w} \propto \bm{\nabla}\rho/\rho$ (see \erefp{eq:thermal_vel}), the ubiquitous, large-amplitude density fluctuations can be interpreted as a manifestation of isotropic thermal velocity.

The perturbed gravitational field induced by FDM density granulation can scatter stars and gas, leaving observable imprints that may help distinguish FDM from other dark matter models. Examples include the outward diffusion of nuclear objects \citep{Chowdhury2021} and stellar streams \citep{Dalal2021}, as well as the dynamical heating of star clusters \citep{Marsh2019}, dwarf galaxies \citep{Dalal2022, Chowdhury2023, Teodori2026, Yang2025b, Yang2025c}, and galactic disks \citep{Church2019, Chiang2023, Yang2024}. These interference effects can also influence dynamical friction \citep{Lancaster2020, Wang2022, Vicente2022, Vitsos2023, Foote2023}, gravitational lensing \citep{Chan2020, Laroche2022, Amruth2023, Powell2023}, and pulsar timing signals \citep{Khmelnitsky2014, DeMartino2017, Blas2017, Smarra2023}. See also \citet{Hui2017, BarOr2019, ElZant2020, Chavanis2021} for further discussions.

\section{Numerical methods}
\label{sec:methods}

This section introduces a variety of FDM algorithms: wave-based schemes (\sref{subsec:methods_wave}), fluid-based schemes (\sref{subsec:methods_fluid}), hybrid approaches (\sref{subsec:methods_hybrid}), adaptive mesh refinement (\sref{subsec:methods_amr}), eigenmode methods (\sref{subsec:methods_eigen}), and collisionless $N$-body methods (\sref{subsec:methods_nbody}).

\subsection{Wave-based methods}
\label{subsec:methods_wave}

The time integration of the Schr\"odinger--Poisson equations can be approximated to second-order accuracy using a split-step method \citep{Taha1984, Woo2009, Mocz2017, Edwards2018, Angulo2022}:
\begin{align}
  \psi(\bm{r},t+\Dt) &= \mathcal{T} \exp\left[ -i \int^{t+\Dt}_t {\rm d}t'
                        \left( -\frac{\hbar}{2m}\nabla^2 + \frac{m}{\hbar}V(\bm{r},t') \right) \right]
                        \psi(\bm{r},t) \\
                     &\approx \mathcal{K}(t+\Dt,\Dt/2)\, \mathcal{D}(t+\Dt/2,\Dt)\, \mathcal{K}(t,\Dt/2)\, \psi(\bm{r},t) + \mathcal{O}(\Dt^3),
  \label{eq:KDK}
\end{align}
where $\mathcal{T}$ is the time ordering operator, and
\begin{align}
  \mathcal{K}(t,\Dt) &= \exp\left[ -i\frac{m}{\hbar}\Dt V(t)\right],
  \label{eq:kick} \\
  \mathcal{D}(t,\Dt) &= \exp\left[ i\frac{\hbar}{2m}\Dt\nabla^2 \right],
  \label{eq:drift}
\end{align}
represent the `kick' and `drift' operators, respectively. This method separates the nonlinear evolution due to the potential term (kick) from the linear evolution due to the kinetic term (drift). For a given wave function and potential at time $t$, the evolution proceeds by first applying the kick operator for a half time step, followed by the drift operator for a full time step. The potential is then recomputed at $t+\Dt$, and a final half-step kick is applied. This procedure is analogous to the `kick-drift-kick (KDK)' scheme commonly employed in $N$-body simulations. To enhance computational efficiency, the initial kick of the current step and the last kick of the previous step can be combined into a single full-step update. Higher-order schemes are also available \citep{Levkov2018, Schwabe2020}.

The kick operator involves only algebraic operations in position space. Specifically, it rotates the phase of the wave function while leaving the density unchanged. The corresponding time-step constraint is commonly set by limiting the phase change per time step to below $2\pi$ \citep{Feit1982}, thereby avoiding phase aliasing:
\begin{equation}
  \Dt_{\mathcal{K}} = \eta_{\mathcal{K}} \frac{2\pi \hbar}{m} \frac{1}{\| V \|_{\rm max}},
  \label{eq:dt_kick}
\end{equation}
where $\eta_{\mathcal{K}} \le 1$ is a safety factor, and $\| V \|_{\rm max}$ denotes the maximum absolute value of the potential. However, since only phase differences are dynamically relevant, it may be more physically motivated to require that the relative phase variation between adjacent simulation elements per time step remain below $\pi$.

Similarly, the drift operator involves only algebraic operations in Fourier space, $\exp[-i\hbar\Dt k^2/2m]$, where $k$ is the wavenumber. Requiring the phase change in the exponent to stay below $2\pi$ over a single time step yields the following time-step criterion:
\begin{equation}
  \Dt_{\mathcal{D}} = \eta_{\mathcal{D}} \frac{4m}{\pi\hbar} \Dh^2,
  \label{eq:dt_drift}
\end{equation}
where the maximum wavenumber in one dimension is taken as $k_{\rm max}=\pi/\Dh$. The safety factor satisfies $\eta_{\mathcal{D}} \le 1$ in general, though its optimal value depends on the adopted wave scheme and specific application. The scaling $\Dt_{\mathcal{D}} \propto \Dh^2$ is characteristic of diffusion-like equations. From the fluid interpretation, $k_{\rm max}$ corresponds to the maximum bulk velocity resolvable at spatial resolution $\Dh$ (see \erefp{eq:bulk_vel}). Accordingly, this time-step condition is analogous to ensuring that a fluid element with the maximum velocity travels no more than one grid cell per time step.

In the following subsections, we introduce various wave schemes for solving the drift operator.

\subsubsection{Global Fourier method}
\label{subsec:methods_global_fourier}

As mentioned above, the linear kinetic term in the Schr\"odinger equation becomes a scalar function in momentum space. Therefore, a pseudo-spectral method is arguably the algorithm of choice for integrating the drift operator. Specifically, it proceeds by first applying a discrete Fourier transform to the wave function, multiplying it by the drift operator in Fourier space, and then applying an inverse discrete Fourier transform to return the updated wave function to position space. This procedure can be schematically expressed as
\begin{equation}
  \mathcal{D}(t,\Dt) = \mathcal{F}^{-1} \exp\left[ -i\frac{\hbar}{2m}\Dt k^2 \right] \mathcal{F},
  \label{eq:drift_fourier}
\end{equation}
where $\mathcal{F}$ and $\mathcal{F}^{-1}$ denote the discrete Fourier transform operator and its inverse, respectively. We refer to this approach as the `global Fourier method' throughout this article to distinguish it from the \emph{local} pseudo-spectral method introduced in \sref{subsec:methods_local_pseudospectral}. Representative codes implementing this method include \textcode{GAMER} \citep{Schive2014a, Kunkel2025}, \textcode{AREPO} \citep[also referred to as \textcode{AxiREPO;}][]{Mocz2017, May2021}, \textcode{PyUltraLight} \citep{Edwards2018} and \textcode{SPoS} built on top of \textcode{Enzo} \citep{Li2019}.

The global Fourier method offers several distinct advantages. The use of the fast Fourier transform (FFT) makes it highly efficient on Central Processing Units (CPUs) and well suited for Graphics Processing Unit (GPU) acceleration. It achieves spectral convergence, with errors decaying faster than any algebraic rate as the number of grid points increases. Consequently, compared to other schemes, the global Fourier method can achieve comparable accuracy using significantly lower spatial resolution and larger time steps, particularly because $\Dt \propto \Dh^2$. Furthermore, the method is unconditionally stable \citep{Taha1984}, which, in principle, allows for using $\eta_{\mathcal{D}}>1$ in \eref{eq:dt_drift}, depending on the accuracy requirements of the application. Finally, the method conserves mass to machine precision.

Despite these promising properties, the global Fourier method has a key limitation. The discrete Fourier transform assumes uniform spatial discretization and periodic boundary conditions. These constraints make the method unsuitable for use at refinement levels in adaptive mesh refinement (AMR) simulations, where compact stencils and Dirichlet or Neumann boundary conditions are typically required on locally refined grids. This limitation motivates the development of alternative approaches, such as finite-difference and local pseudo-spectral methods discussed in the following subsections.

\subsubsection{Finite-difference methods}
\label{subsec:methods_fd}

Finite-difference methods are well-suited for refined regions in AMR simulations and naturally support aperiodic boundary conditions. For example, the \textcode{AxioNyx} code \citep{Schwabe2020} employs a fourth-order finite-difference method. The \textcode{GAMER} \citep{Schive2014a} and \textcode{SCALAR} \citep{Mina2020} codes expand the drift operator in position space as
\begin{equation}
  \mathcal{D}(t,\Dt) = \sum_{p=0}^{q} \frac{1}{p\,!}\left[ i\frac{\hbar}{2m}\Dt\nabla^2\right]^p.
  \label{eq:drift_fd}
\end{equation}
The von Neumann stability analysis \citep{NumericalRecipes} shows that this scheme is conditionally stable for $q\ge3$. See also \sref{subsec:methods_finite_volume} for a finite-volume method implemented in the \textcode{GIZMO} code \citep{Hopkins2019}.

The one-dimensional scheme along the $x$-direction with $q=3$ is expressed as
\begin{equation}
  \mathcal{D}(t,\Dt) =  1 + i\left(\frac{\hbar}{2m}\Dt\right)\frac{\partial^2}{\partial x^2}
                          - \frac{1}{2}\left(\frac{\hbar}{2m}\Dt\right)^2\frac{\partial^4}{\partial x^4}
                          - \frac{i}{6} \left(\frac{\hbar}{2m}\Dt\right)^3\frac{\partial^6}{\partial x^6}.
  \label{eq:drift_fd_3rd}
\end{equation}
The Laplacian operator can be approximated using a second-order finite-difference scheme:
\begin{equation}
  \frac{\partial^2}{\partial x^2}\psi^n_j = \frac{ \psi^n_{j+1} - 2\psi^n_j + \psi^n_{j-1} }{\Dh^2},
  \label{eq:laplacian_2nd}
\end{equation}
where $n$ and $j$ denote the time step and cell index, respectively. The resulting Courant--Friedrichs--Lewy (CFL) stability condition is
\begin{equation}
  \Dt_{\mathcal{D}} = \frac{\sqrt{3}m}{2\hbar}\Dh^2,
  \label{eq:dt_drift_fd}
\end{equation}
corresponding to $\eta_{\mathcal{D}} \approx 0.7$. Adopting a higher-order approximation for the Laplacian operator can reduce numerical errors, at the cost of larger stencils, increased computational cost, and a slightly more stringent CFL condition. Extension to three dimensions is straightforward using dimensional splitting, $\exp[\nabla^2] = \exp[\partial^2_z]\,\exp[\partial^2_y]\,\exp[\partial^2_x]$, where the order of operations is arbitrary since derivatives along different directions commute.

Finite-difference methods do not, in general, conserve mass, where the total mass is defined as $M_{\rm tot}=\sum_j\|\psi_j\|^2$. This issue can be resolved by additionally solving the continuity equation, \eref{eq:madelung_continuity}. Here we describe a straightforward one-dimensional implementation as an illustration. First, the continuity equation can be discretized as
\begin{equation}
  \rho^{n+1}_j = \rho^n_j - \frac{\Dt}{\Dh}\left( F^{n+1/2}_{j+1/2} - F^{n+1/2}_{j-1/2} \right),
  \label{eq:fd_mass_continuity}
\end{equation}
where $F^{n+1/2}_{j\pm1/2}$ are the mass fluxes at the right and left interfaces of cell $j$ at the half-step time $t+\Dt/2$. To estimate the half-step fluxes, we note that the finite-difference scheme in \eref{eq:drift_fd_3rd} is equivalent to the following two-step formulation:
\begin{align}
  \psi^{n+1/2}_{j} &= \left[ 1 + \frac{i}{2}\left(\frac{\hbar}{2m}\Dt\right)\frac{\partial^2}{\partial x^2}
                               - \frac{1}{6}\left(\frac{\hbar}{2m}\Dt\right)^2\frac{\partial^4}{\partial x^4} \right]\psi^{n}_{j},
  \label{eq:drift_fd_3rd_half} \\
   \psi^{n+1}_{j}  &= \psi^{n}_{j} + \left[ i\left(\frac{\hbar}{2m}\Dt\right)\frac{\partial^2}{\partial x^2} \right]\psi^{n+1/2}_{j}.
  \label{eq:drift_fd_3rd_full}
\end{align}
Accordingly, we use the intermediate wave function $\psi^{n+1/2}_{j}$ from \eref{eq:drift_fd_3rd_half} to compute $F^{n+1/2}_{j\pm1/2}$ using Eqs. (\ref{eq:mass_dens}--\ref{eq:bulk_vel}):
\begin{align}
  F^{n+1/2}_{j+1/2} &= \frac{\hbar}{m}\left(  R^{n+1/2}_{j+1/2}\frac{\partial I^{n+1/2}_{j+1/2}}{\partial x}
                                            - I^{n+1/2}_{j+1/2}\frac{\partial R^{n+1/2}_{j+1/2}}{\partial x}  \right),
  \label{eq:fd_mass_flux} \\
  R^{n+1/2}_{j+1/2} &= \frac{R^{n+1/2}_{j+1} + R^{n+1/2}_{j}}{2},
  \label{eq:fd_mass_real} \\
  \frac{\partial R^{n+1/2}_{j+1/2}}{\partial x} &= \frac{R^{n+1/2}_{j+1} - R^{n+1/2}_{j}}{\Dh},
  \label{eq:fd_mass_grad}
\end{align}
where $I^{n+1/2}_{j+1/2}$ and $\partial I^{n+1/2}_{j+1/2}/\partial x$ are defined analogously. Finally, we rescale the full-step solution $\psi^{n+1}_{j}$ from \eref{eq:drift_fd_3rd_full} to match the mass-conserving density $\rho^{n+1}_j$ from \eref{eq:fd_mass_continuity}:
\begin{equation}
  \psi^{n+1}_{j,\rm rescaled} = \psi^{n+1}_{j}\frac{\sqrt{\rho^{n+1}_j}}{\|\psi^{n+1}_{j}\|}.
  \label{eq:fd_mass_rescale}
\end{equation}

Importantly, this fluid-based correction remains valid even in the presence of vortices, since the mass flux (Eq. \ref{eq:fd_mass_flux}) is well-behaved near nodes (see \sref{subsec:challenges_vortices}). Moreover, for AMR simulations, $F^{n+1/2}_{j\pm1/2}$ can be reused to correct inter-level fluxes at resolution boundaries to preserve mass conservation. This capability is implemented in both \textcode{GAMER} and \textcode{SCALAR}. One caveat, however, is that this hybrid approach may degrade the error convergence rate and alter the stability condition of the original finite-difference methods, particularly when higher-order Laplacian approximations are employed.

The major drawback of finite-difference methods lies in its significantly slower error convergence compared to the global Fourier method (e.g., see \fref{fig:test_gaussian}). As a result, in AMR simulations employing the global Fourier method at the root level and a finite-difference method at refinement levels, it is typically necessary to activate multiple refinement levels simultaneously within a given region to surpass the accuracy of a root-level-only simulation (see \sref{subsec:methods_amr} for further discussion). This limitation motivates the development of local pseudo-spectral methods, which use finite stencils like finite-difference methods but achieve comparable accuracy at lower resolution, as discussed in the next subsection.

\subsubsection{Local pseudo-spectral methods}
\label{subsec:methods_local_pseudospectral}

The global Fourier method is limited to periodic domains, while finite-difference methods exhibit much slower convergence. In addition, AMR simulations favor compact stencils to minimize computational and communication overhead. To meet these requirements, a variety of algorithms have been proposed to achieve fast convergence on non-periodic uniform grids. Here, we briefly introduce a local pseudo-spectral method based on Fourier continuations with Gram polynomials \citep{Lyon2009, Lyon2010}, hereafter referred to as the `FC--Gram' method. This method has been implemented in \textcode{GAMER} \citep{Kunkel2025}. It provides an efficient, accurate, and stable approach to extend non-periodic data onto a periodic grid, thereby enabling the use of a pseudo-spectral Fourier method and achieving high-order algebraic convergence.

\begin{figure*}[th!]
\includegraphics[width=\textwidth]{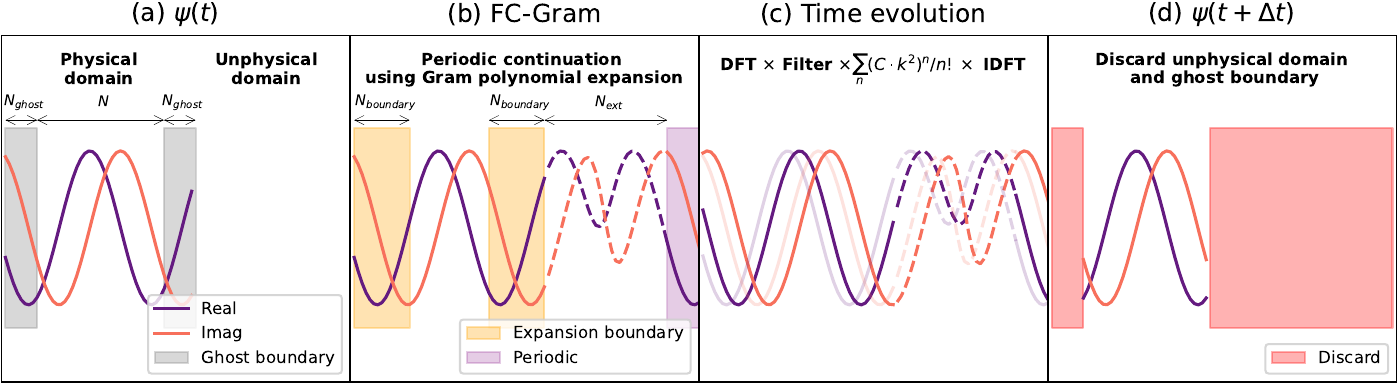}
\caption{
Overview of the local pseudo-spectral FC--Gram method. Solid and dashed lines represent the wave function in the physical and periodically extended regions, respectively. The method consists of four main steps. (a) Filling ghost cells at the domain boundaries. (b) Extending the non-periodic data onto a periodic grid using Fourier continuations with Gram polynomials. (c) Evolving the wave function in the periodic domain using the discrete Fourier transform (DFT) with a spectral filter, where $C=-i\hbar\Dt/2m$. (d) Discarding data in the ghost cells and the extended region. See text for details.
Image reproduced with permission from \citet{Kunkel2025}, copyright by AAS
}
\label{fig:FC_Gram_solver}
\end{figure*}

\fref{fig:FC_Gram_solver} provides an overview of the FC--Gram method in one dimension, which consists of four main steps:
\begin{enumerate}[label=(\alph*)]
  \item Filling ghost cells at the domain boundaries.
  \item Extending the non-periodic data onto a periodic grid.
  \item Evolving the wave function in the periodic domain.
  \item Discarding data in the ghost cells and the extended region.
\end{enumerate}

In Step (a), ghost-cell data of size $N_{\rm ghost}$ are added to each side of a target grid of size $N$. These ghost cells can be filled by applying physical boundary conditions, copying from adjacent grids at the same resolution, or, in AMR simulations, interpolating from nearby coarse grids. This yields a physical domain of size $N+2N_{\rm ghost}$. Importantly, ghost cells serve a dual purpose. In addition to enabling data exchange among neighboring grids, they also help prevent unphysical data in the periodically extended region from contaminating the target grid, provided that an appropriate time-step constraint is imposed.

Step (b) is the most critical. It introduces smooth, periodic continuations of the non-periodic data to avoid the Gibbs phenomenon and enable the use of rapidly convergent Fourier methods. Specifically, it extends the non-periodic wave function from the physical domain into an additional region of size $N_{\rm ext}$, thereby forming a \emph{periodic} computational domain of size $N_{\rm total}=N+2N_{\rm ghost}+N_{\rm ext}$. The wave function in this extended region is \emph{unphysical} in the sense that it does not represent real simulation data. Instead, it is artificially constructed solely to create a smooth periodic connection between the two ends of the physical domain.

To generate this extension, the method first projects the wave function from $N_{\rm boundary}$ cells on each side of the physical domain onto Gram polynomials. These discrete orthogonal polynomials are then replaced with a set of Fourier modes that both match the Gram polynomials within the physical domain and remain periodic over the full $N_{\rm total}$ domain. The Fourier coefficients are determined by solving a linear optimization problem using singular value decomposition, which minimizes the mismatch between the Gram polynomials and their Fourier continuations within the physical domain. See \citet{Kunkel2025} for details.

In Step (c), the wave function in the full computational domain of size $N_{\rm total}$ is updated by a time step $\Dt$. Since this extended domain is periodic, the wave function can be evolved straightforwardly using a pseudo-spectral Fourier method, similar to the scheme described in \sref{subsec:methods_global_fourier}. There is, however, an important caveat: the artificial periodic extension can introduce high-$k$ modes that lead to numerical instability and pollute the wave function in the physical domain. This issue can be mitigated by truncating the Taylor expansion of the drift operator in Fourier space, $\exp[-i\hbar\Dt k^2/2m]$, to order $N_{\rm Taylor}$ and further applying an exponential filter:
\begin{equation}
  \exp\left(-\mu (k/k_{\rm max})^{2\nu}\right) \sum_{p=0}^{N_{\rm Taylor}} \frac{(-i\hbar\Dt k^2/2m)^p}{p!},
  \label{eq:FC_Gram_filter}
\end{equation}
where the filter parameters are set to $\mu=16\ln(10)$, $\nu=50$, and $k_{\rm max} = \pi/\Dh$ by default \citep{Albin2011}. This combination of Taylor truncation and exponential filtering ensures numerical stability by suppressing spurious high-$k$ modes.

Finally, in Step (d), only the data within the original grid of size $N$ are retained.

The error convergence rate of the FC--Gram method depends on the order of the Gram polynomials. With the default configuration in \textcode{GAMER}, the method achieves eleventh-order accuracy. It thus significantly outperforms finite-difference methods, although it still falls short of the global Fourier method (see \fref{fig:test_gaussian}). Mass conservation, however, is not guaranteed to machine precision. This is because the pseudo-spectral Fourier update in Step (c) is unitary only on the full computational domain, with no guarantee of a discrete local conservation law. As a result, the effective inter-grid mass fluxes computed by adjacent grids generally do not match exactly. The empirically determined CFL condition is $\eta_{\mathcal{D}} \approx 0.2\text{--}0.3$. This time-step criterion must ensure not only numerical stability but also that the unphysical wave function in the periodically extended region does not propagate into the physical domain beyond the ghost boundaries. Finally, note that Steps (b)--(d) are all linear operations and can be consolidated into a single matrix multiplication, thereby improving computational efficiency for small grids.

\subsection{Fluid-based methods}
\label{subsec:methods_fluid}

A key limitation in wave-based methods is the need to resolve the de Broglie wavelength even where the density is smooth and nonzero (see \sref{subsec:challenges_wavelength}). Moreover, these methods do not in general guarantee conservation of mass, momentum, and energy, although mass conservation can be restored by additionally evolving the continuity equation. These limitations motivate fluid-based approaches, solving either the Hamilton--Jacobi--Madelung equations, Eqs. (\ref{eq:hjm_continuity}--\ref{eq:hjm_phase}), or the Madelung equations, Eqs. (\ref{eq:madelung_continuity}--\ref{eq:madelung_momentum}). Fluid methods have several notable advantages over wave approaches. They require much lower resolution in regions with smooth density and velocity, enabling significantly larger simulation volumes. They ensure manifest conservation when solving the fluid equations in conservative forms. They can readily incorporate a Lagrangian formulation, thereby respecting Galilean invariance and supporting adaptive resolution. However, fluid approaches have inherent limitations---they cannot accurately resolve strong interference, since the velocity and quantum potential diverge at density nodes (see \sref{subsec:challenges_vortices}).

In the following, we introduce two families of fluid-based schemes: smoothed particle hydrodynamics (\sref{subsec:methods_sph}) and finite-volume/finite-difference methods (\sref{subsec:methods_finite_volume}).

\subsubsection{Smoothed particle hydrodynamics}
\label{subsec:methods_sph}

Smoothed particle hydrodynamics (SPH) is widely used in astrophysical simulations \citep{Monaghan1992, Springel2010}. In SPH, the fluid is represented by Lagrangian particles. The physical quantity $F$ at the position of particle $a$, $F_a = F(\bm{r}_a)$, is approximated as a kernel-weighted sum over neighbors: $F_a=\sum_b(m_b/\rho_b)F_bW_{ab}(h)$, where $m_b$ and $\rho_b$ are the mass and density of neighbor $b$, and $W_{ab}(h) = W(\|\bm{r}_a - \bm{r}_b\|, h)$ is a smooth, spherically symmetric kernel function. Here, $h$ is the kernel (smoothing) length, which generally varies in space and time to keep the number of neighbors roughly constant. Given the particle-sampled fluid fields and their derivatives, the system is advanced in time by solving the hydrodynamic equations in Lagrangian form.

SPH appears readily applicable to FDM simulations by replacing the isotropic gas-pressure term with the quantum acceleration $-\bm{\nabla}Q$ in \eref{eq:madelung_velocity}. Accurately evaluating $\bm{\nabla}Q$ is, however, highly nontrivial because it involves third derivatives of the density field, which are prone to noise and numerical instability. \citet{Nori2018} proposed the following formulation implemented in the \textcode{AX-GADGET} code built on top of \textcode{P-GADGET3}:
\begin{equation}
  \frac{{\rm d}^2\bm{r}_a}{{\rm d}t^2}
    = -\bm{\nabla} Q_a
    = \frac{\hbar^2}{2m^2} \sum_b \frac{m_b}{f_b\rho_b}
      \left[ \frac{\nabla^2\rho_b}{2\rho_b} - \frac{\|\bm{\nabla}\rho_b\|^2}{4\rho_b^2} \right]
      \bm{\nabla} W_{ab},
  \label{eq:sph_acc}
\end{equation}
where
\begin{equation}
  f_b = 1 + \frac{h_b}{3\rho_b}\frac{\partial \rho_b}{\partial h_b}
  \label{eq:sph_corr}
\end{equation}
is a correction accounting for variable smoothing lengths and ${\rm d}/{\rm d}t$ is the convective derivative. \eref{eq:sph_acc} follows from applying the variational principle to the discretized SPH Lagrangian, with Lagrange multipliers enforcing the constraint $\rho_bh_b^3 = {\rm const}$. The gradient and Laplacian of the density field are computed as
\begin{align}
  \bm{\nabla}\rho_a &= \sum_b (\rho_b - \rho_a) \frac{m_b\bm{\nabla} W_{ab}}{\sqrt{\rho_a \rho_b}},
  \label{eq:sph_gra_rho} \\
  \nabla^2\rho_a    &= \sum_b (\rho_b - \rho_a) \frac{m_b\nabla^2 W_{ab}}{\sqrt{\rho_a \rho_b}}
                       - \frac{\|\bm{\nabla}\rho_a\|^2}{\rho_a}.
  \label{eq:sph_lap_rho}
\end{align}
For other SPH-like formulations, see \citet{Mocz2015, Veltmaat2016, Zhang2018a}.

In principle, in addition to the conventional hydrodynamic time-step constraints, a diffusion-like criterion, $\Delta t \propto h^2$ (similar to \erefp{eq:dt_drift}), is required for numerical stability and accuracy. However, this condition has not been enforced in the aforementioned SPH codes; its impact thus remains to be assessed.

SPH offers several benefits owing to its Lagrangian nature. The method supports smoothly adaptive resolution and can be easily coupled to $N$-body gravity solvers. It is Galilean invariant, which minimizes advection errors relative to Eulerian methods. However, it faces several significant challenges. The quantum potential $Q$ depends on higher-order spatial derivatives of the density, which can be very noisy with standard SPH derivative estimators. Both $Q$ and velocity diverge at vortices where $\rho \rightarrow 0$ (see \sref{subsec:challenges_vortices} and \fref{fig:test_halo_slice}), which can cause substantial truncation errors, hinder numerical convergence, and force prohibitively small time steps. The smoothing length, generally scaling as $h \propto \rho^{-1/3}$ (for fixed particle mass), can exceed the local de Broglie wavelength $\LdB$ in low-density regions, making it difficult to resolve density granulation and the associated quantum pressure within halos.

These limitations likely contribute to the following discrepancies between different SPH implementations and wave-based methods. First, it remains uncertain whether SPH can robustly resolve solitons and recover the soliton--halo relation in FDM halos. For example, the flat cores found in the SPH spherical-collapse and cosmological simulations of \citet[][Fig. 4]{Zhang2018a}, \citet[][Fig. 4]{Nori2021}, and \citet[][Fig. 7]{Nori2023} do not exhibit the sharp soliton--halo transition commonly predicted in wave-based simulations (e.g., see \fref{fig:test_halo_profile}) and observed in the SPH soliton-collision simulations of \citet[][Fig. 4]{Veltmaat2016}. Relatedly, the velocity dispersion within the flat cores in \citet[][Fig. 5]{Zhang2018a} and \citet[][Fig. 8]{Nori2023} remains roughly constant. This is inconsistent with the soliton solution, which is supported by quantum pressure rather than velocity dispersion, as confirmed by wave-based simulations \citep[e.g.,][]{Liao2025}. Furthermore, \citet{Nori2018} found that the matter power spectrum is suppressed on small scales by the quantum pressure relative to $N$-body simulations. This contrasts with the results of the SPH simulations by \citet{Veltmaat2016} and the wave-based global spectral simulations by \citet{May2021}, which reported enhanced small-scale power at low redshifts due to wave interference. These limitations and controversies highlight ongoing uncertainties and raise concerns about the reliability of SPH simulations in capturing the coarse-grained properties of FDM.

\subsubsection{Finite-volume and finite-difference methods}
\label{subsec:methods_finite_volume}

\citet{Hopkins2019} presented a family of Lagrangian, meshless finite-volume and finite-mass Godunov schemes implemented in the \textcode{GIZMO} code to solve the Madelung equations, Eqs. (\ref{eq:madelung_continuity}) and (\ref{eq:madelung_momentum}). The simulation volume is partitioned into moving, unstructured meshes defined by a weighted kernel, resembling a Voronoi tessellation but with smoothed mesh boundaries. Each mesh carries volume-averaged conserved quantities, such as mass and momentum density, and moves with an arbitrary velocity $v_{\rm mesh}$. The conserved quantities are updated using a finite-volume formulation by estimating fluxes across mesh boundaries. If one sets $v_{\rm mesh}$ equal to the local dark matter bulk velocity $v$, the inter-mesh mass fluxes vanish, and the mass of each mesh remains constant during evolution. It thus eliminates the need to solve the continuity equation. This approach is known as the meshless finite-mass method.

The code employs a second-order, matrix-based gradient estimator, which is essential for accurately computing the quantum stress tensor that involves higher-order derivatives of the mass density. Time integration is performed using an explicit leapfrog scheme with adaptive time steps. In addition to the conventional CFL conditions in hydrodynamics, a time-step stability criterion similar to \eref{eq:dt_drift} is employed, with the grid spacing $\Dh$ replaced by the kernel length $h$. Gravity is incorporated via operator splitting. Inter-mesh fluxes are computed using an HLL-type Riemann solver in the rest frame of the mesh interface, with the left and right states reconstructed using a piecewise-constant method. This scheme conserves mass and momentum to machine precision. To further ensure energy conservation, the method introduces an auxiliary scalar field that tracks the energy associated with `unresolved' quantum potential arising from numerical dissipation. This energy is then coupled back into the momentum equation as an additional isotropic pressure term.

The meshless finite-volume/mass formulation provides several notable advantages. It shares the key properties of Lagrangian methods, as in SPH. It conserves mass, momentum, and (optionally) energy, even with adaptive time steps. Moreover, it is significantly more stable and accurate than SPH methods, especially at `well-behaved' nodes where the velocity remains finite while the density approaches zero. This improvement is partly because the formulation evolves conserved quantities and the quantum stress tensor, both of which do not diverge at nodes (see \fref{fig:test_halo_slice} for an illustration). However, this approach still faces a common limitation of all fluid-based methods: difficulty handling `divergent’ nodes, such as vortices, where the velocity becomes singular (see \sref{subsec:challenges_vortices} for further discussion). In addition, similar to SPH, its Lagrangian nature tends to concentrate resolution in high-density regions, making it difficult to resolve halo granules when $\LdB < h$. Consequently, it remains unclear how this approach performs relative to other fluid- and wave-based methods in more realistic, complex settings---for instance, whether it can resolve fine-grained wave structure or preserve coarse-grained FDM features in cosmological simulations.

A similar finite-volume method can be applied to evolve the wave function by noting that the Schr\"odinger equation can also be written in a flux-conservative form:
\begin{align}
  & \frac{\partial R}{\partial t} + \bm{\nabla}\cdot \left(\frac{ \hbar}{2m}\nabla I \right) = \frac{ m}{\hbar} V I,
  \label{eq:schroedinger_real} \\
  & \frac{\partial I}{\partial t} + \bm{\nabla}\cdot \left(-\frac{\hbar}{2m}\nabla R \right) = -\frac{m}{\hbar} V R,
  \label{eq:schroedinger_imag}
\end{align}
with gravity treated as a source term. However, this approach does not guarantee conservation of mass, momentum, or energy, due to their nonlinear dependence on the wave function. Mass conservation can be recovered by additionally solving the continuity equation and rescaling the wave function's amplitude while keeping its phase unchanged, similar to the approach outlined in \sref{subsec:methods_fd}.

\citet{Kunkel2025} proposed a finite-difference method to solve the Hamilton--Jacobi--Madelung equations, Eqs. (\ref{eq:hjm_continuity}--\ref{eq:hjm_phase}), on the fluid levels of a hybrid scheme (see \sref{subsec:methods_hybrid}). The continuity equation is solved using a monotonic upstream-centered scheme for conservation laws (MUSCL), combined with a piecewise-linear data reconstruction and the van Albada limiter \citep{Albada1997}. For the phase equation, the advection term $\|\bm{\nabla} S\|^2$ is computed using the Sethian--Osher flux \citep{Osher1988}, and the quantum potential term $\nabla^2 \sqrt{\rho}/\sqrt{\rho}$ is discretized using the second-order central finite-difference method. Time integration is performed using a strong stability-preserving, third-order Runge--Kutta method \citep{Shu2007}. The time-step criteria depend on both \eref{eq:dt_drift} and the spatial derivative of the phase field. This method conserves mass but not momentum or energy.

\subsection{Hybrid methods}
\label{subsec:methods_hybrid}

Fluid-based schemes are significantly more efficient than wave-based schemes for addressing smooth, high-velocity flows, which occupy most of the simulation volume outside collapsed regions in cosmological simulations. Nevertheless, fluid schemes cannot robustly resolve strong interference in multi-stream regions, where wave schemes are preferred. This dilemma motivates hybrid approaches, which apply fluid schemes in smooth, single-stream regions and switch to wave schemes in interference-dominated, multi-stream regions. In addition, hybrid schemes facilitate cosmological zoom-in simulations, as purely wave-based schemes cannot accurately capture the large-scale structure when the de Broglie wavelength outside the zoom-in regions is poorly resolved. The Lagrangian zoom-in regions can be identified either by performing collisionless $N$-body simulations or by introducing tracer particles into FDM simulations, analogous to Bohmian mechanics \citep{Wyatt2005}.

One critical challenge in such hybrid approaches is handling the fluid--wave interfaces---specifically, how to define a unique and accurate mapping between fluid variables and the wave function to impose consistent boundary conditions on both sides. Consequently, in fluid regions, it is generally preferable to evolve the phase field via the Hamilton--Jacobi--Madelung equations, Eqs. (\ref{eq:hjm_continuity}--\ref{eq:hjm_phase}), rather than evolving the velocity field via the Madelung equations, Eqs. (\ref{eq:madelung_continuity}--\ref{eq:madelung_velocity}), since converting velocity to wave function is non-unique and subject to a time-dependent phase factor. The initial phase field can be computed by solving \eref{eq:cosmo_ic_phase} given the velocity distribution.

\citet{Veltmaat2018} implemented a hybrid $N$-body--wave method in the AMR code \textcode{Enzo}. On coarser AMR levels (treated as fluid regions), the quantum pressure is ignored and the dynamics are approximated by a collisionless $N$-body approach. On the finest level, the code evolves the wave function using a finite-difference method. To reconstruct the wave function from $N$-body particles, a `classical wave function' formulation is adopted, in which each particle represents a localized wave packet with a central phase $S$. The phase of particle $a$ is updated by solving \eref{eq:hjm_phase} in Lagrangian form without the quantum potential $Q$:
\begin{equation}
  \frac{{\rm d}S_a}{{\rm d}t} = \frac{m}{2\hbar}v_a^2 - \frac{m}{\hbar} V(\bm{r}_a),
  \label{eq:hjm_phase_convective}
\end{equation}
where ${\rm d}/{\rm d}t = \partial/\partial t + (\hbar/m)\bm{\nabla}S \cdot \bm{\nabla}$ is the convective derivative. The phase of the wave function is reconstructed by superposing wave packets carried by particles using an interpolation kernel. The amplitude is obtained by applying the same kernel to particle mass (thus SPH-like) rather than to wave packets, thereby ensuring mass conservation but omitting interference effects. This reconstruction serves two purposes: (i) providing initial conditions for the wave solver in a target Lagrangian volume prior to shell-crossing; (ii) supplying boundary conditions at the fluid--wave interfaces for subsequent wave evolution. Conversely, no phase extraction from the wave function is required, because \eref{eq:hjm_phase_convective} involves no spatial derivatives of $S$.

One key limitation in this classical wave approximation is its inability to capture interference fringes. As a result, the initial wave function (on the highest AMR level) must be reconstructed in single-stream regions, and the fluid--wave interfaces must lie well outside the halo virial radius. To eliminate this restriction, \citet{Schwabe2022} proposed reconstructing the full wave function, including both phase and amplitude, from wave packets via a WKB-like Gaussian beam method with fixed amplitudes:
\begin{equation}
  \psi(\bm{r},t) = \sum_a W(\|\bm{r}-\bm{r}_a\|,h) \exp{\left[ i\left( S_a + \frac{m}{\hbar} \bm{v}_a \cdot (\bm{r}-\bm{r}_a) \right) \right]},
  \label{eq:gaussian_beam}
\end{equation}
where $W$ is a Gaussian kernel with smoothing length $h$. This improved method captures statistically correct interference patterns, enabling wave function reconstruction even in multi-stream regions well within the halo virial radius. However, it introduces mass conservation errors, which are mitigated by rescaling the average density of the wave function to match the underlying total particle mass. In addition, a time-step constraint is imposed to limit the maximum phase change per step in \eref{eq:hjm_phase_convective} to less than $\pi$. This hybrid method is implemented in the AMR code \textcode{AxioNyx}.

\begin{figure*}[th!]
\includegraphics[width=\textwidth]{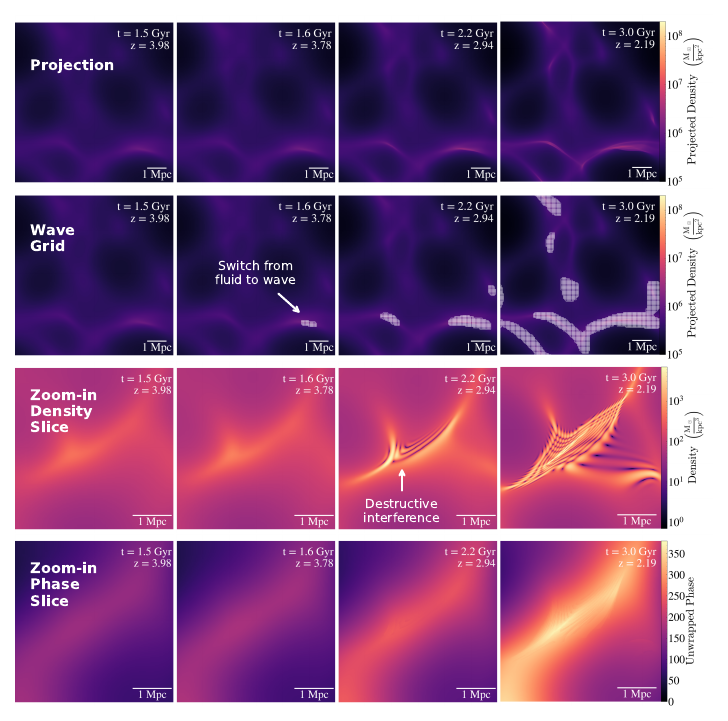}
\caption{
An FDM cosmological simulation using the hybrid fluid--wave scheme of \citet{Kunkel2025}. The wave solver is activated at redshift $z = 3.83$, prior to the development of interference around $z = 3.1$. The top two rows show the projected density, with white grids indicating regions evolved by the wave solver. The bottom two rows present the zoom-in views of the density and unwrapped phase slices centered on the first wave region.
Image reproduced with permission from \citet{Kunkel2025}, copyright by AAS
}
\label{fig:hybrid_scheme}
\end{figure*}

In these $N$-body--wave methods, the particles' classical trajectories neglect the effects of quantum potential, and the reconstructed wave function is only accurate on a statistical level. To improve upon these limitations, \citet{Kunkel2025} proposed a fully grid-based hybrid scheme in the AMR code \textcode{GAMER}. On coarser levels, a finite-difference fluid scheme solving the Hamilton--Jacobi--Madelung equations in Eulerian form is employed (see \sref{subsec:methods_finite_volume}). It retains the quantum potential term and thus remains accurate until divergent velocity and phase discontinuities develop at density voids. The phase difference between adjacent cells can substantially exceed $2\pi$ in general, reflecting the fact that fluid-based schemes do not need to resolve the de Broglie wavelength. Finer levels evolve the wave function using the local pseudo-spectral FC--Gram method described in \sref{subsec:methods_local_pseudospectral}. See \fref{fig:hybrid_scheme} for an illustration.

Reconstructing the wave function from the fluid fields is unambiguous and straightforward, as both the density/phase and wave function are defined on grids, thus eliminating the need for an interpolation kernel. However, unlike in $N$-body--wave methods, the fluid scheme here also requires boundary conditions at the fluid--wave interfaces, since \eref{eq:hjm_phase} involves spatial derivatives of both $S$ and $\rho$. A complication arises from the fact that the phase extracted from the wave function is only defined up to a multiple of $2\pi$. To resolve this ambiguity, the fine-grid phase in the wave region is unwrapped to align smoothly with the coarse-grid phase in the fluid region. This phase unwrapping implicitly assumes that the de Broglie wavelength is resolved by at least two grid points across the fluid--wave interfaces, which necessitates sufficiently fine resolution before switching to the wave formulation. While this may appear to increase computational cost, it is in fact a prerequisite for any wave-based schemes to avoid significant aliasing errors. See \sref{subsec:methods_amr} for the corresponding grid refinement criteria.

\subsection{Adaptive mesh refinement}
\label{subsec:methods_amr}

Uniform-resolution methods are inefficient for cosmological simulations due to the large dynamic range involved. This motivates the application of AMR to FDM \citep{Schive2014a, Schwabe2020, Mina2020, Kunkel2025}, which allows spatial and temporal resolution to adjust dynamically based on prescribed refinement criteria. However, designing robust and efficient refinement criteria for FDM simulations is nontrivial. Conventional quasi-Lagrangian or super-Lagrangian refinement strategies are insufficient, since the de Broglie wavelength (and thus the characteristic granule size) is generally independent of density. Velocity-based criteria are problematic near vortices, where the velocity diverges.

One common strategy is to apply the L\"ohner error estimator \citep{Lohner1987}, based on the ratio of the second to the first derivatives, to both the real and imaginary parts of the wave function. In the hybrid scheme of \citet{Kunkel2025}, a spectral refinement criterion based on the decay rate of the polynomial coefficients in the FC--Gram method (see \sref{subsec:methods_local_pseudospectral}) is applied to the wave levels. On the fluid levels, refinement is determined by the quantum potential and the Laplacian of the phase to detect strong interference and phase discontinuities.

Data exchange between different refinement levels involves interpolating coarse-grid data to fine grids and averaging fine-grid data to coarse grids, often referred to as the `prolongation' and `restriction' operations, respectively. Applying these operations directly to the wave function does not ensure mass conservation. This issue can be addressed by also interpolating/averaging the density field and then rescaling the amplitude of the wave function while preserving its phase, similar to the density correction method described in \sref{subsec:methods_fd}. Furthermore, for smooth, high-velocity flows, commonly found outside halos, interpolating/averaging the smooth density and phase fields gives more accurate results than operating directly on the rapidly oscillating wave function. However, the phase is discontinuous near vortices, where phase interpolation fails. Accordingly, \citet{Kunkel2025} proposed a hybrid interpolation scheme that switches between density/phase and wave-function interpolation based on the local second derivative of the phase field. Conservation errors can also arise at coarse--fine resolution interfaces, which, when solving the conservative fluid formulation, can be corrected by reconciling mismatches between fine- and coarse-grid fluxes, as commonly applied in hydrodynamic AMR simulations.

AMR supports adaptive time steps: lower-resolution levels can take larger time steps to reduce computational cost, which is particularly useful in FDM simulations due to the $\Dt \propto \Dh^2$ scaling. However, it introduces an underappreciated issue in wave-based, self-gravitating simulations---the boundary values of gravitational potential used on leaf patches for the Poisson solver can be incorrect. The reason is that these boundary values are obtained by interpolating the coarse-grid potential. Under adaptive time steps, coarse and fine grids are generally asynchronous, which necessitates temporal interpolation. This, in turn, requires advancing the density on coarser (non-leaf) patches before leaf patches, which is problematic for wave-based methods because effective mass fluxes can be highly inaccurate when the de Broglie wavelength is under-resolved. Applying the density correction scheme introduced in \sref{subsec:methods_fd} does not resolve this issue, as the mass fluxes---and thus the resulting coarse-grained mass \emph{distribution}---remain inaccurate, even though the total mass is conserved. One possible improvement is to compute coarse-grid fluxes by averaging those computed on leaf patches, rather than deriving them from the coarse-grid wave function. Another potential solution is to adopt a momentum-conserving wave-based scheme, which, to our knowledge, has yet to be developed. Note that this issue does not arise in conventional hydrodynamic AMR simulations, for which finite-volume methods ensure manifest momentum conservation regardless of resolution.

\subsection{Eigenmode methods}
\label{subsec:methods_eigen}

Fully self-consistent Schr\"odinger--Poisson solvers can be computationally prohibitive, especially for large $\mFDM$. Furthermore, when studying the properties and impacts of FDM interference in isolated, nearly static, dark-matter-dominated systems, it is often reasonable to neglect the gravitational backreaction from non-dark-matter components. These considerations motivate the use of eigenmode-based methods, which approximate the time-dependent wave function as a linear superposition of eigenmodes of a \emph{static}, smooth gravitational potential $V_0(\bm{r})$, thereby enabling efficient modeling of FDM halos that incorporate interference substructures.

The method involves three main steps:
\begin{enumerate}[label=(\alph*)]
  \item For a target smooth density distribution $\rho_0(\bm{r})$, compute the corresponding gravitational potential $V_0(\bm{r})$.
  \item Compute a library of bound-state eigenmodes associated with $V_0(\bm{r})$.
  \item Determine a suitable superposition of eigenmodes such that the time-averaged density distribution best approximates $\rho_0(\bm{r})$ and remains stable.
\end{enumerate}
These steps can be iterated, using the output density from Step (c) as the updated input $\rho_0(\bm{r})$ for Step (a), until convergence is achieved.

The expansion in Step (c) reads
\begin{equation}
  \psi(\bm{r},t) = \sum_j a_j \psi_j(\bm{r}) e^{-iE_jt/\hbar},
  \label{eq:eigen_expansion}
\end{equation}
where $\psi_j$ is the $j\text{-th}$ eigenmode with energy eigenvalue $E_j$, and $a_j$ is the corresponding time-independent complex amplitude with a random phase. For a spherically symmetric potential $V_0(r)$, one commonly adopts $\psi_j=R_{n\ell}(r)Y_\ell^m(\theta,\phi)$, where $R_{n\ell}$ is the radial function, $Y_\ell^m$ are the spherical harmonics, and $n,\ell,m$ are the radial, angular momentum, and magnetic quantum numbers, respectively. The corresponding energy eigenvalue  $E_j=E_{n\ell}$ is independent of $m$ due to spherical symmetry. Crucially, rather than numerically solving Eqs. (\ref{eq:schroedinger}--\ref{eq:poisson}), the wave function can be evolved analytically by propagating the phase of each eigenmode via $\exp(-iE_jt/\hbar)$, offering significant computational savings.

The density field is given by
\begin{align}
  \rho(\bm{r},t) &= \|\psi(\bm{r},t)\|^2 \\
                 &= \sum_j \|a_j\|^2 \|\psi_j(\bm{r})\|^2
                    + \sum_{j \ne k} a_j a_k^* \psi_j(\bm{r}) \psi_k^*(\bm{r}) e^{i(E_k-E_j)t/\hbar}.
  \label{eq:eigen_density}
\end{align}
The first term on the right-hand side represents the time-independent average profile, whereas the second term describes the time-dependent interference patterns (i.e., density granulation), which statistically average out over time due to the random phases. The task in Step (c) is therefore to determine the coefficients $a_j$ such that the sum $\sum_j \|a_j\|^2 \|\psi_j(\bm{r})\|^2$ matches the target smooth density $\rho_0(\bm{r})$ as closely as possible for self-consistency. Below, we briefly review several proposed methods to achieve this.

\citet{Lin2018} adopted the fermionic King model \citep{Chavanis1998} to assign the coefficients $a_j$. This approach arises from comparing several theoretical distribution functions with $a_j$ directly measured from halos in FDM cosmological simulations. The discrepancy between the density profiles in Steps (a) and (c) is corrected iteratively by treating the potential difference as a perturbed Hamiltonian. The shape of the central core (i.e., the soliton) is determined by the ground-state solution, while its amplitude is set by the soliton--halo relation \citep{Schive2014b}. One limitation of this method is the inability to specify arbitrary density profiles, as the final converged profile may differ noticeably from the initial guess. Furthermore, since the adopted theoretical distribution function is calibrated using simulation halos in a limited halo mass range, $2.8\times10^9\Msun\text{--}7.0\times10^{10}\Msun$, and at a fixed FDM particle mass, $\mFDM=0.81$, it is questionable whether the method remains reliable beyond this range. These concerns motivate the development of the alternative methods discussed below.

\citet{Dalal2021} adopted the Widrow--Kaiser ansatz \citep{Widrow1993}, which constructs the wave function as a superposition of plane waves weighted by a classical distribution function $f(\bm{r},\bm{p})$:
\begin{equation}
  \psi(\bm{r}) = m^{1/2} \sum_{\bm{p}} \left[ f(\bm{r},\bm{p}) d\bm{p} \right]^{1/2} N_{\bm{p}} e^{i\bm{r} \cdot \bm{p}/\hbar}.
  \label{eq:eigen_wk_ansatz}
\end{equation}
Here, $d\bm{p}$ is the discrete momentum-space resolution, and $N_{\bm{p}}$ are complex random numbers satisfying $\ave{ N_{\bm{p}}^*N_{\bm{q}} } = \delta_{\bm{pq}}$, where $\ave{}$ denotes the ensemble average over different realizations. The distribution function $f(\bm{r},\bm{p})$ must satisfy the self-consistency condition: $m \sum_{\bm{p}} f(\bm{r},\bm{p}) d\bm{p} = \rho_0(\bm{r})$. Under the assumptions of spherical symmetry and isotropy, $f(\bm{r},\bm{p})$ can be derived from the Eddington formula \citep{Binney2008, Teodori2026}. To evolve the wave function using \eref{eq:eigen_expansion}, the coefficients $a_j$ can be obtained by projecting the wave function from \eref{eq:eigen_wk_ansatz} onto the eigenmodes $\psi_j$. Unlike the method of \citet{Lin2018}, this approach supports more general density profiles. One caveat, however, is that $f(\bm{r},\bm{p})$ cannot accurately model the central soliton core. This limitation may be addressed by superposing a theoretical soliton wave function following the soliton--halo relation. Finally, we note that \eref{eq:eigen_wk_ansatz} can also be used to construct a uniform FDM background with density granulation by inserting a Maxwell-Boltzmann distribution into $f(\bm{r},\bm{p})$ \citep{Lancaster2020}.

\citet{Yavetz2022} generalized the classical, particle-based Schwarzschild method \citep{Schwarzschild1979} to construct FDM halos by replacing particle orbits with wave eigenmodes \citep[see also][]{Yang2025a, Zimmermann2025}. The coefficients $a_j$ are determined by formulating an optimization problem that minimizes the difference between the output density from Step (c) and the input $\rho_0(\bm{r})$ in Step (a). Additional constraints can be imposed during the optimization---for example, requiring $a_j$ to depend only on energy, analogous to an isotropic distribution function in particle-based models. A random phase is assigned to each $a_j$, mirroring the random orbital phases in the original Schwarzschild method. In the halo outskirts, where the Wentzel–Kramers–Brillouin (WKB) approximation holds, this approach is consistent with the Widrow--Kaiser ansatz. In contrast, in the central region, where the WKB-limit breaks down and the distribution function derived from the Eddington formula becomes invalid, the Schwarzschild method remains applicable and can support rather arbitrary cored profiles, which need not resemble conventional soliton-like solutions.

Compared to Schr\"odinger--Poisson solvers, eigenmode methods are significantly more computationally efficient. They also allow for simulating a restricted region of the halo without concern for boundary effects. However, evolving the wave function via \eref{eq:eigen_expansion} is only approximate, as it neglects the nonlinear interactions between eigenmodes and the gravitational backreaction from external objects.

This class of methods has been applied to a variety of studies, including the temporal and spatial correlation of density granules \citep{Lin2018}, soliton random walk, oscillations, and distortions \citep{Li2021, Zagorac2022}, and granulation-induced dynamical heating of stellar streams \citep{Dalal2021} and ultra-faint dwarf galaxies \citep{Dalal2022}. In addition, eigenmode methods can be used solely to generate stable FDM initial conditions for Schr\"odinger--Poisson solvers. Example applications include studies of dynamical friction \citep{Lancaster2020}, soliton--halo relation \citep{Yavetz2022}, and granular heating of galactic disks \citep{Yang2024} and dwarf spheroidal galaxies \citep{Teodori2026}.

\subsection{Collisionless $N$-body methods}
\label{subsec:methods_nbody}

The conventional collisionless $N$-body methods offer a convenient and efficient way to capture the large-scale dynamics of FDM, as the Schr\"odinger--Poisson equations reduce to the Vlasov--Poisson equations in the limit $\LdB \to 0$ \citep{Widrow1993}. In practice, the simulation setup follows that of CDM simulations, except that the initial density power spectrum is generated using the FDM transfer function (Eqs. \ref{eq:transfer_func1} and \ref{eq:transf_func2}), making it analogous to warm dark matter (WDM) simulations. Compared to fully wave-based FDM simulations, this $N$-body approach allows for much larger comoving volumes due to its numerical efficiency. However, by neglecting quantum pressure, it fails to capture fine-grained wave phenomena, such as density granulation and soliton cores.

The $N$-body methods, though approximate, have a number of practical applications. They can probe the FDM halo mass function in large-scale structure \citep{Schive2016, May2022}, as the suppression of low-mass FDM halos is mainly determined by the simulation initial conditions (see \fref{fig:cdm_vs_fdm}). They are also useful for studying the morphology of FDM halos \citep{Dome2023}. By comparing approximate $N$-body and genuine FDM simulations with the same initial conditions, one can quantify the dynamical effects of quantum pressure \citep{Veltmaat2016, Nori2018, May2021} and assess the numerical convergence of genuine FDM simulations \citep[][see also \fref{fig:test_cosmo_profile}]{Liao2025}. In hybrid fluid--wave methods, the $N$-body approaches can provide the large-scale gravitational field and the boundary conditions for wave regions (see \sref{subsec:methods_hybrid}), and can also be used to identify the Lagrangian volumes of selected halos in cosmological zoom-in simulations \citep{Veltmaat2018, Chan2025}.

Several studies have attempted to constrain the FDM particle mass using $N$-body simulations, such as through the Lyman-alpha forest power spectrum \citep[e.g.,][]{Irsic2017} and the subhalo mass function \citep[e.g.,][]{Nadler2024}. However, it remains unclear whether omitting strong interference fringes and quantum pressure effects---such as their impact on the tidal evolution of subhalos and solitons \citep{Chan2025}---may bias these constraints. Moreover, $N$-body approaches are susceptible to spurious halo formation, albeit with properties differing quantitatively from those in genuine FDM simulations (see \fref{fig:spurious_halo}).

\section{Numerical challenges}
\label{sec:challenges}

This section examines key numerical challenges in FDM simulations: the small time and length scales associated with the de Broglie wavelength (\sref{subsec:challenges_wavelength}), the singular behavior at vortices (\sref{subsec:challenges_vortices}), and the formation of spurious halos (\sref{subsec:challenges_spurious_halos}).

\subsection{de Broglie wavelength}
\label{subsec:challenges_wavelength}

\begin{figure*}[th!]
\includegraphics[width=\textwidth]{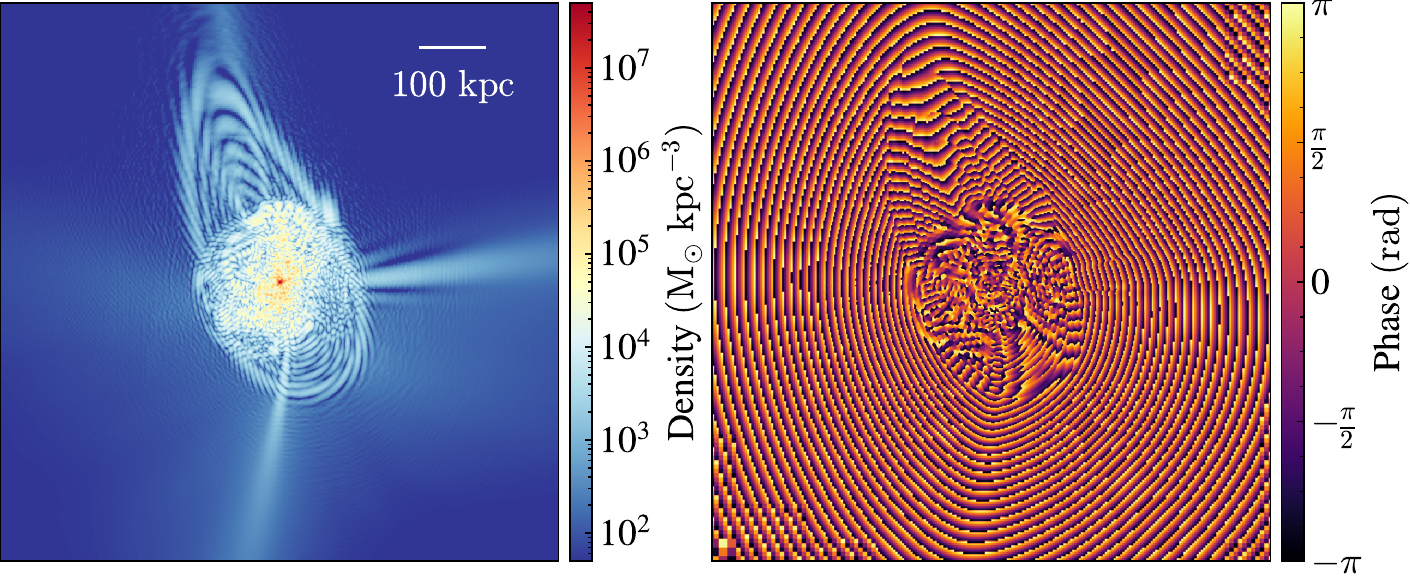}
\caption{
Slices of the density (left) and phase (right) fields through the center of a $\Mh=4.5\times10^{11}\Msun$ halo at $z \sim 0.8$ with $\mFDM=0.1$. Outside the halo, the density remains smooth while the phase exhibits rapid oscillations, corresponding to plane-wave solutions of the wave function. This requires wave-based schemes to employ much higher resolution than fluid-based schemes in order to accurately resolve the phase structure. Inside the halo, strong interference leads to ubiquitous, large-amplitude density fluctuations, necessitating high resolution for both wave-based and fluid-based schemes to resolve the de Broglie wavelength
}
\label{fig:phase}
\end{figure*}

The necessity of resolving the short de Broglie wavelength and its rapid oscillations associated with high-velocity flows poses a significant challenge for FDM simulations, especially for large $\mFDM$. The manifestation of the de Broglie wavelength in FDM cosmological simulations can be broadly categorized into two regimes: (i) plane waves in low-density regions outside halos, and (ii) strong interference fringes within halos and along filaments. See \fref{fig:phase} for an illustration.

The large-scale, high-velocity flows in the free-fall regions outside halos correspond to plane-wave solutions of the wave function, $\psi \propto \exp(i\bm{k}\cdot\bm{r}-\omega t)$, with a wavelength $\LdB = h/mv \approx 1.2 \,\mFDM^{-1}(v/100\kms)^{-1} \kpc$. As a result, unlike CDM simulations, which typically employ high resolution only in dense regions, FDM cosmological simulations require high resolution even in low-density regions, where the density is smooth but the velocity is high. Failure to resolve the de Broglie wavelength associated with these high-velocity flows can lead to an underestimation of the velocity field, resulting in delayed structure formation. These low-density regions occupy a substantial fraction of the simulation volume, which is the primary reason why wave-based FDM simulations reaching lower redshifts are typically limited to comoving box sizes of only a few Mpc. By contrast, FDM simulations using fluid-based schemes can largely overcome this limitation, since the density, velocity, and phase fields are all smooth outside halos, allowing coarser resolution without resolving the de Broglie wavelength.

Strong interference leads to widespread, isotropic density granulation within virialized halos, as well as long, thin fringes along filaments. In contrast to plane waves, which exhibit oscillating phase but smooth density, the density field in these interference regions also displays large-amplitude fluctuations on the scale of the de Broglie wavelength associated with the local velocity dispersion (see \erefp{eq:granule_size}). As a result, both fluid-based and wave-based schemes require similarly high spatial resolution. For example, for a $\Mh \sim 10^{12} \Msun$ halo with $\vdisp \sim 100 \kms$ and $\mFDM=10$, one would need a minimum spatial resolution of at least $\sim 10 \pc$. Insufficient resolution can lead to inaccurate estimation of the quantum potential and quantum pressure, both of which involve the second spatial derivative of the density (see Eqs. \ref{eq:quantum_potential} and \ref{eq:quantum_pressure}), potentially resulting in unphysical halo contraction or expansion and inaccurate soliton properties \citep{Liao2025}. Furthermore, density voids generated by strong destructive interference are ubiquitous within halos and can lead to the formation of vortices. In these zero-density regions, both the velocity and quantum potential diverge, making fluid-based schemes particularly vulnerable to large numerical errors. See \sref{subsec:challenges_vortices} for further discussion.

Virialized FDM halos are approximately isothermal. For example, for a halo of mass $\sim 10^{12}\Msun$, the temperature in the central dense region is only about a factor of 2 higher than that near the virial radius, despite a density contrast of several orders of magnitude \citep[e.g., see][]{Liao2025}. This near-isothermality implies that consistently high spatial resolution is required throughout the entire halo, resulting in substantial computational cost. Moreover, this requirement for nearly uniform resolution poses a particular challenge for Lagrangian approaches, such as SPH and moving mesh methods, whose spatial resolution typically scales with local mass density as $\rho^{-1/3}$. As a result, these methods may lack sufficient resolution in the low-density outskirts of halos, making it difficult to resolve density granules and, in turn, the quantum pressure in those regions.

The requirement of high spatial resolution for resolving density granules also necessitates high temporal resolution to capture their rapid oscillations, as $T_{\rm gra} \propto \mFDM d_{\rm gra}^2$ (see \erefp{eq:granule_time}). This constraint applies equally to both fluid and wave schemes.

The computational costs as functions of halo mass $\Mh$ and FDM particle mass $\mFDM$ can be estimated as follows. The characteristic velocity dispersion scales as $\vdisp \propto (\Mh/\rh)^{1/2} \propto \Mh^{1/3}$, where $\rh \propto \Mh^{1/3}$ is the halo virial radius. The spatial and temporal resolution required to resolve density granulation, $\Dh$ and $\Dt$, thus scale as
\begin{align}
  \Dh &\propto (\mFDM \vdisp)^{-1} \propto \mFDM^{-1} \Mh^{-1/3},
  \label{eq:sim_cost_dh} \\
  \Dt &\propto \mFDM \Dh^2 \propto \mFDM^{-1} \Mh^{-2/3}.
  \label{eq:sim_cost_dt}
\end{align}
We assume here that the simulation time steps are dominated by the drift operator, \eref{eq:dt_drift}, which generally holds within halos. The memory consumption, $C_{\rm mem}$, is proportional to the total number of simulation elements $N \propto (\rh/\Dh)^3$. The simulation time, $C_{\rm time}$, scales as $N/\Dt$, assuming that each time step requires $\mathcal{O}(N)$ operations rather than, e.g., $\mathcal{O}(N \ln N)$. These yield
\begin{align}
  C_{\rm mem} &\propto \mFDM^3\Mh^2,
  \label{eq:sim_cost_memory} \\
  C_{\rm time} &\propto \mFDM^4\Mh^{8/3}.
  \label{eq:sim_cost_time}
\end{align}
The strong dependence on $\mFDM$ and $\Mh$ implies that computational costs can quickly become computationally prohibitive for large $\mFDM$ and $\Mh$.

\subsection{Vortices}
\label{subsec:challenges_vortices}

\begin{figure*}[th!]
\centering
\includegraphics[width=0.5\textwidth]{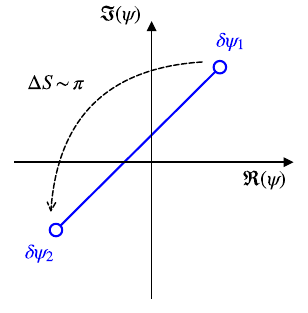}
\caption{
An illustration of the numerical challenge in resolving vortices within regions of vanishing density. The two open circles represent the simulation elements. Although the infinitesimal wave function $\delta \psi$ varies smoothly near the vortex at the origin, the phase field $S$ exhibits rapid variation, and the corresponding velocity diverges. This steep phase gradient and divergent velocity pose significant challenges for fluid-based schemes
}
\label{fig:phase_jump}
\end{figure*}

Strong interference in an FDM halo can generate zero-density regions. They appear as closed loops located at the intersection of the two isosurfaces $\Re({\psi})=0$ and $\Im(\psi)=0$, where $\Re({\psi})$ and $\Im(\psi)$ denote the real and imaginary parts of the wave function, respectively. Along these nodal lines, the phase field $S$ becomes multi-valued, rendering the velocity field $\bm{v} \propto \bm{\nabla}S$ ill-defined. As a result, the flow vorticity $\bm{\nabla} \times \bm{v}$ is a Dirac delta function, and the velocity circulation $\oint \bm{v} \cdot d\bm{l}$ enclosing such a nodal line is quantized due to phase winding. These closed loops of zero density can thus be interpreted as vortex rings \citep{Chiueh2011, Hui2021a}. The number density of vortex rings is approximately one per de Broglie volume, implying that vortices are ubiquitously distributed throughout each halo.

In the vicinity of a vortex line, the wave function remains well-behaved, but the phase field exhibits rapid variation. See \fref{fig:phase_jump} (and also \fref{fig:test_vortex_pair}) for an illustration. The density and velocity fields scale as $\rho \sim r^2$ and $v \sim r^{-1}$, respectively, where $r$ is the distance to the vortex line. Consequently, both the velocity and the quantum potential ($Q \propto {\nabla}^2 \sqrt{\rho}/\sqrt{\rho}$) diverge as $r \rightarrow 0$ (see \fref{fig:test_halo_slice}).

\begin{figure*}[th!]
\includegraphics[width=\textwidth]{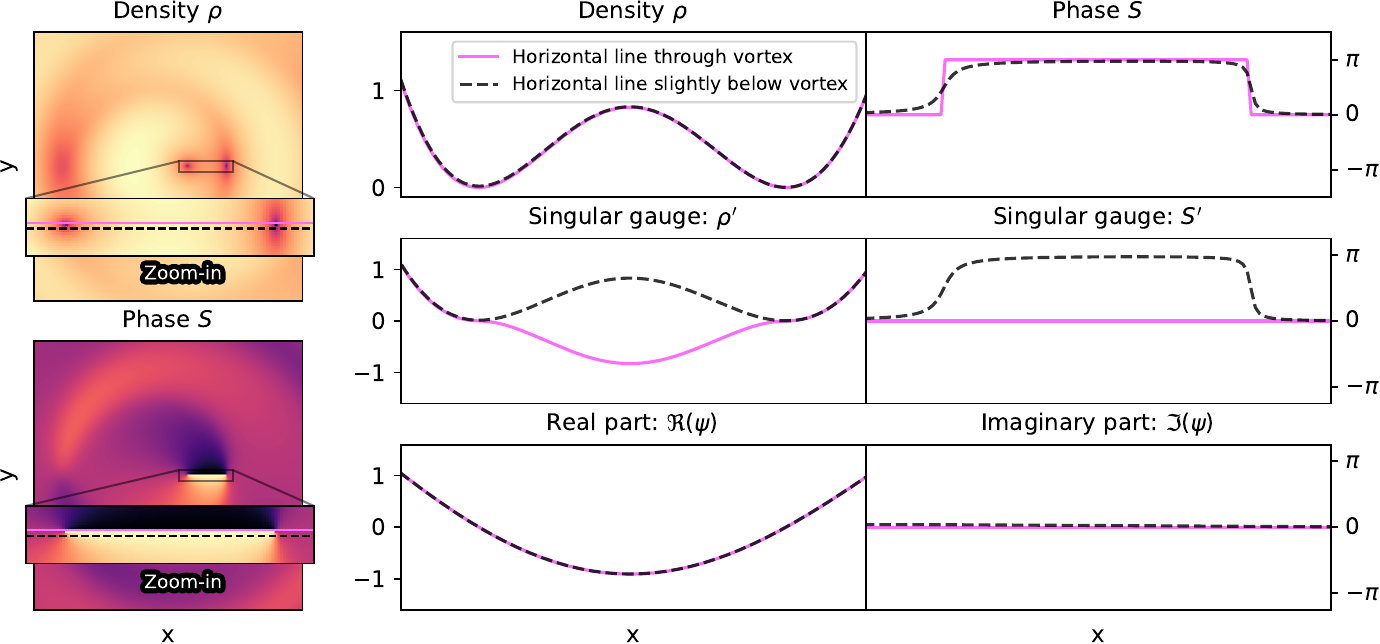}
\caption{
Wave function near a corotating vortex pair described by \eref{eq:vortex_pair} at $t=0$. The two panels on the left present slices of density and phase, with insets showing a close-up of the vortices, which feature zero density and discontinuous phase. The panels on the right display profiles along lines directly through (solid) and slightly below (dashed) the vortex pair (see insets on the left). While the density and wave function vary smoothly across a vortex, the phase field exhibits a $\pi$ jump exactly at the vortex and changes rapidly nearby (see also \fref{fig:phase_jump}). This steep phase gradient cannot be alleviated by the singular gauge transformation, $\rho' = -\rho$ and $S' = S + \pi$.
Image reproduced with permission from \citet{Kunkel2025}, copyright by AAS
}
\label{fig:vortex_phase}
\end{figure*}

As an example, \fref{fig:vortex_phase} shows the distribution of wave function, phase, and density near a vortex pair described by \eref{eq:vortex_pair} at $t=0$. The wave-based schemes work well for vortices since both $\Re({\psi})$ and $\Im(\psi)$ vary smoothly. However, the discontinuity in the phase field is numerically problematic for simulations evolving the Hamilton--Jacobi--Madelung equations, Eqs. (\ref{eq:hjm_continuity}--\ref{eq:hjm_phase}). Although applying a singular gauge transformation, $\rho' = -\rho$ and $S' = S + \pi$, can regularize both the density and phase exactly at the vortex, the phase still varies rapidly in its vicinity. Increasing spatial resolution offers limited improvement, if any, because the phase gradient becomes even steeper closer to the vortex, as $\|\bm{\nabla}S\| \propto r^{-1}$. Similarly, the divergence of the velocity and quantum potential poses significant challenges for simulations based on the Madelung equations, Eqs. (\ref{eq:madelung_continuity}--\ref{eq:madelung_velocity}). A further complication arises for Lagrangian approaches, whose inherently lower resolution in underdense regions makes it difficult to resolve vortices accurately.

Solving the momentum conservation equation, \eref{eq:madelung_momentum}, may seem promising at first glance, since at vortices the momentum density $\rho v$ vanishes, and both the momentum density flux $\rho v^2$ and the stress tensor $\Sigma_{ij} \propto \rho \frac{\partial^2 \ln\rho}{\partial x_i \partial x_j}$ remain finite (as shown in \fref{fig:test_halo_slice}). However, computing the momentum density flux involves evaluating the ratio of two infinitesimal quantities, $(\rho v)^2$ and $\rho$. Likewise, an accurate estimation of the stress tensor depends on a delicate cancellation between the divergent term $\frac{\partial^2 \ln\rho}{\partial x_i \partial x_j}$ and the vanishing term $\rho$. These calculations are therefore highly susceptible to large numerical errors.

In short, resolving vortices using fluid-based schemes or Lagrangian approaches requires the development of innovative algorithms. It remains unclear whether, and to what extent, the inability to capture local vortices would affect the intrinsic global properties of FDM halos, such as the granular halo structure and its associated dynamical heating effects, the amplitude of quantum pressure, the degree of energy equipartition between kinetic and thermal energies, and the soliton--halo relation.

\subsection{Spurious halos}
\label{subsec:challenges_spurious_halos}

It is well established that WDM simulations suffer from the formation of low-mass spurious halos, resulting in an unphysical upturn at the low-mass end of the halo mass function \citep{Wang2007, Angulo2013, Schneider2013}. These spurious halos arise from artificial fragmentation along filaments and are seeded by numerical artifacts. Consequently, their masses and positions are sensitive to numerical accuracy, and simulations with different resolutions generally yield distinct distributions of spurious halos. In addition, their protohalos at the initial redshift of a simulation tend to exhibit flattened, disk-like geometries. In comparison, the distribution of genuine halos shows good convergence with increasing simulation accuracy, and their protohalos are typically more spheroidal \citep{Lovell2014}.

\begin{figure*}[th!]
\includegraphics[width=\textwidth]{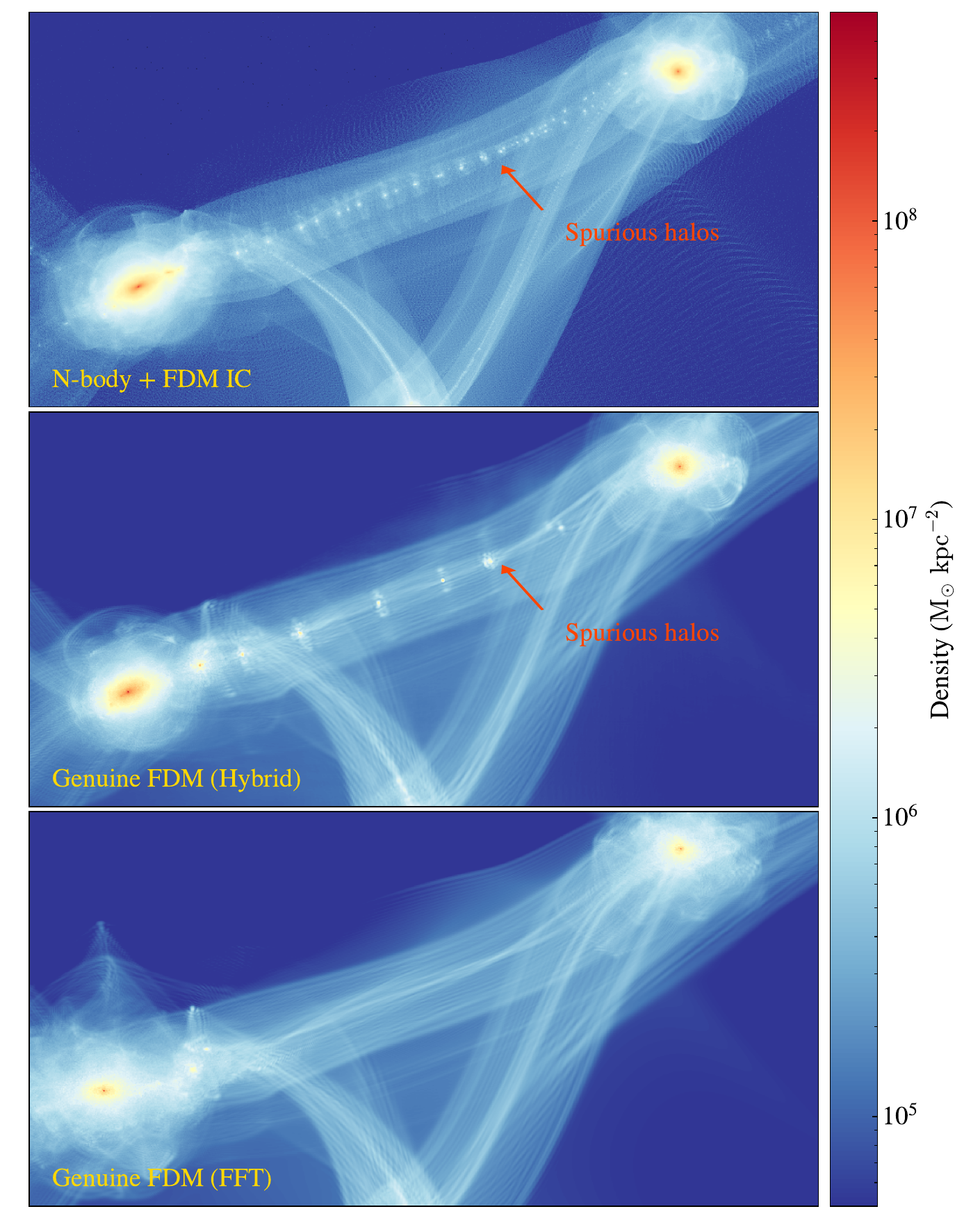}
\caption{
An illustration of spurious halos in FDM simulations. The image compares the density distributions along a filament at $z=0$ across three different simulation setups from the same FDM initial condition: (top) a collisionless $N$-body simulation, (middle) a genuine FDM simulation with AMR using a hybrid fluid--wave algorithm, and (bottom) a genuine FDM simulation using the global Fourier method. Spurious halos, highlighted by arrows in the top and middle panels, appear along the filament as regularly spaced clumps with masses substantially below the half-mode mass, $\Mhf \propto \mFDM^{-4/3}$. Notably, these halos lack consistent counterparts in simulations with different setups, as they arise from artificial fragmentation seeded by numerical noise. The bottom panel shows no evidence of spurious halos, presumably due to the superior accuracy of the global Fourier method
}
\label{fig:spurious_halo}
\end{figure*}

Similar to WDM simulations, spurious halos with mass significantly below the half-mode mass, $\Mhf \propto \mFDM^{-4/3}$ (see \erefp{eq:half_mode}), can also form in FDM simulations, especially at lower redshifts. One may wonder whether quantum pressure could alleviate this issue, given that it can suppress small-scale structure. However, the dynamical impact of quantum pressure during the simulation is largely confined to highly nonlinear regions with large-amplitude density fluctuations, such as within granular halos. This aligns with the fact that the suppression of low-mass FDM halos is mainly governed by the sharp cutoff in the linear density power spectrum established at matter-radiation equality (see \sref{subsec:small_scale_suppression}). In filaments, velocity dispersion and quantum pressure are significant only in directions perpendicular to the filament axis and are negligible along it, as indicated by the interference fringes oriented \emph{parallel} to filaments (see the bottom panel in \fref{fig:spurious_halo}). As a result, small-scale perturbations can become Jeans unstable along filaments, but these are primarily seeded by numerical noise rather than physical perturbations. This ultimately leads to the formation of spurious FDM halos.

\fref{fig:spurious_halo} illustrates this issue by comparing the density distributions along a filament across three different simulation setups, all starting from the same FDM initial condition with $\mFDM=0.8$: (i) a collisionless $N$-body simulation (\sref{subsec:methods_nbody}), (ii) a genuine FDM simulation using the hybrid fluid--wave algorithm based on the local pseudo-spectral FC--Gram method (\sref{subsec:methods_hybrid}), and (iii) a genuine FDM simulation using the global Fourier method (\sref{subsec:methods_global_fourier}). The two massive halos at the ends of the filament are presumably genuine, as their masses are comparable to $\Mhf$ and they appear consistently in all three simulations. By contrast, the low-mass clumps along the filament in the upper and middle panels are likely artifacts, as they exhibit several features commonly associated with spurious halos. First, their masses are well below $\Mhf$. Second, their spatial distribution shows a suspiciously regular spacing along the filament. Third, these halos lack consistent counterparts in simulations with different setups, suggesting that they are sensitive to numerical accuracy.

Spurious halos in the $N$-body simulation are expected, since this setup is closely analogous to a WDM simulation, differing only in the exact shape of the initial density power spectrum. Their emergence in the genuine FDM simulation (middle panel) is also not surprising, given the above discussion, although this issue has received comparatively less attention than in the WDM case. For example, do spurious protohalos in FDM simulations also exhibit flatter geometries than genuine halos? How do their characteristic mass and spacing depend on resolution and algorithm? Notably, \fref{fig:spurious_halo} suggests that the global Fourier method may be free from artificial fragmentation, possibly due to its superior accuracy along smooth filaments. However, this method cannot capture fine structures within the two massive halos once the de Broglie wavelength approaches the grid scale.

\section{Test bench}
\label{sec:test}

This section describes a set of representative numerical tests for both validating numerical accuracy and illustrating FDM features. These include Gaussian wave packets (\sref{subsec:tests_gaussian}), vortices (\sref{subsec:tests_vortices}), the Jeans instability (\sref{subsec:tests_jeans}), solitons (\sref{subsec:tests_soliton}), isolated halos (\sref{subsec:tests_halo}), and cosmological simulations (\sref{subsec:tests_cosmo}). Other illustrative tests in the literature include oblique traveling waves \citep{Hopkins2019}, the node problem in plane-wave interference \citep{Hopkins2019}, simple harmonic oscillators \citep{Mocz2015}, the self-similar solution of a density jump \citep{Hui2017, Li2019}, spherical collapse \citep{Schwabe2020}, and two-stream collisions \citep{Hui2017}.

\subsection{Tests without self-gravity}
\label{subsec:tests_no_gravity}

\subsubsection{Gaussian wave packets}
\label{subsec:tests_gaussian}

The Gaussian wave packet provides a simple analytical solution to the free-particle Schr\"odinger equation. The wave function with mean momentum $p_0=mv_0$ and initial mean position $x_0$ is given by
\begin{equation}
  \psi(x,t) = \frac{A}{\sqrt{\Delta^2+\frac{i\hbar}{m}t}}
               \exp{ \left[ -\frac{(x-x_0-v_0 t)^2}{2(\Delta^2+\frac{i\hbar}{m}t)} \right] }
               \exp{ \left[ i\frac{mv_0}{\hbar}\left( x-x_0-\frac{v_0}{2}t \right) \right] },
  \label{eq:gaussian_psi}
\end{equation}
where $\Delta$ represents the position uncertainty and $A = \Delta^{1/2} \pi^{-1/4}$ normalizes the total mass to unity. The corresponding mass density and bulk velocity, derived from Eqs. (\ref{eq:mass_dens}--\ref{eq:bulk_vel}), are
\begin{align}
  \rho(x,t) &= \frac{A^2}{\sqrt{\Delta^2\gamma(t)}} \exp{ \left[ -\frac{(x-x_0-v_0 t)^2}{\gamma(t)} \right] },
  \label{eq:gaussian_rho} \\
  v(x,t) &= v_0 + \frac{(x-x_0-v_0 t)\hbar^2}{\Delta^2 m^2\gamma(t)}t,
  \label{eq:gaussian_vel} \\
  \gamma(t) &= \Delta^2 + \frac{\hbar^2}{\Delta^2m^2}t^2.
  \label{eq:gaussian_gamma}
\end{align}

Unlike a single plane wave, the Gaussian wave packet presents a nontrivial task for fluid-based schemes. It therefore serves as a useful benchmark for comparing wave and fluid schemes and for calibrating their respective optimal resolutions as functions of $v_0$ and $\Delta$. In particular, fluid schemes generally require much lower resolution when evolving a single wave packet with high $v_0$. However, in the case of two colliding wave packets, strong interference can lead to divergent velocity and quantum potential at nodes of zero density, posing numerical challenges for fluid schemes even at very high resolution \citep{Li2019}.

\begin{figure*}[th!]
\centering
\includegraphics[width=\textwidth]{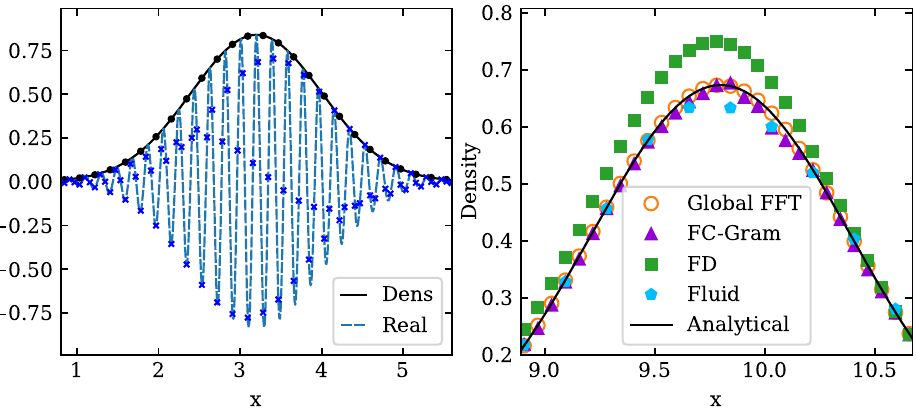}
\caption{
Test of a traveling Gaussian wave packet. (Left) Initial density and real part of the wave function. Crosses and circles represent simulation data from the wave and fluid schemes, with resolutions of $\Dh \approx \LdB/3$ and $\LdB$, respectively. (Right) Simulation results at $t=0.2$ using different numerical schemes. The global Fourier method (circles) yields the most accurate result; the FC--Gram method (triangles) introduces mild distortion, while the wave-based finite-difference method (squares) shows significant deviation from the analytical solution (solid line). Notably, the fluid scheme (pentagons) successfully captures the overall wave packet motion, despite not resolving the de Broglie wavelength ($\Dh \sim \LdB$)
}
\label{fig:test_gaussian}
\end{figure*}

\fref{fig:test_gaussian} shows the evolution of a single traveling Gaussian wave packet with $\Delta = 0.8$, $v_0 = 33$, $x_0 = 3.2$, $A = 0.67$, and $m/\hbar = 1$. The left panel displays the initial condition at $t=0$. The de Broglie wavelength associated with the bulk velocity, $\LdB=2\pi/v_0$, is much shorter than the Gaussian width, illustrating the advantage of fluid-based methods in this regime. The right panel compares numerical results at $t=0.2$ obtained using four different schemes: the global Fourier method, the local pseudo-spectral FC--Gram method, the wave-based finite-difference method, and the fluid-based finite-difference method solving the Hamilton--Jacobi--Madelung equation (see \sref{sec:methods} for details). To test the robustness of each approach, a deliberately low resolution of approximately three cells per $\LdB$ is adopted for the wave methods. The fluid method, by contrast, uses a resolution three times lower and therefore does not resolve $\LdB$. No AMR is applied. As shown, the global Fourier method closely reproduces the analytical solution. The FC--Gram method introduces mild distortion, while the wave-based finite-difference method exhibits significant deviations. Notably, the fluid method performs well despite its substantially coarser resolution.

\subsubsection{Vortices}
\label{subsec:tests_vortices}

As discussed in \sref{subsec:challenges_vortices}, vortices are ubiquitous in FDM halos, where the phase field is discontinuous and both the velocity and quantum potential diverge. As a result, vortices pose a serious challenge for fluid-based methods. In contrast, wave-based methods remain robust, as the wave function varies smoothly across vortices. In hybrid schemes, vortices can serve as benchmarks for calibrating the criteria of switching between fluid and wave methods and for monitoring potential numerical artifacts at the fluid--wave interfaces.

\begin{figure*}[th!]
\includegraphics[width=\textwidth]{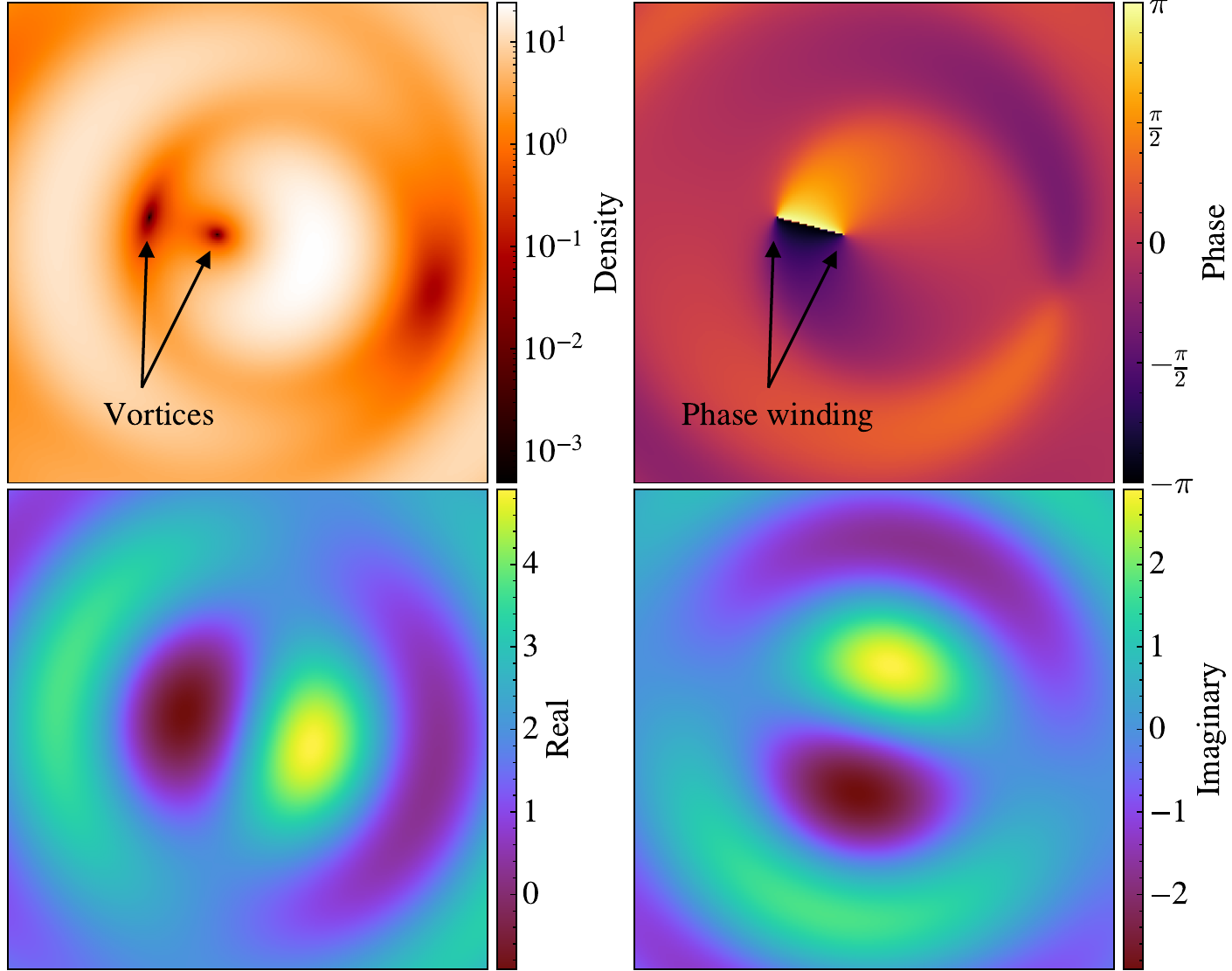}
\caption{
A corotating vortex pair described by \eref{eq:vortex_pair}. See text for details
}
\label{fig:test_vortex_pair}
\end{figure*}

As an example, \eref{eq:vortex_pair} describes a corotating vortex pair:
\begin{equation}
  \psi(R, \varphi) = \rho_0 - B {\rm J_1}\left(R\sqrt{\frac{2 \omega m}{\hbar}}\right) {\rm e}^{i (\varphi - \omega t)},
  \label{eq:vortex_pair}
\end{equation}
where ${\rm J_1}$ is the Bessel function of the first kind, $\omega$ is the angular frequency, $\rho_0$ is the background density, $B$ is a constant, $R$ is the cylindrical radius, and $\varphi$ is the polar angle. \fref{fig:test_vortex_pair} shows the analytical results for $\rho_0=2$, $B=5$, $\omega=90$, and $m/\hbar=1$ at $t=0.8$. Note that the $2\pi$ phase jump perpendicular to the line connecting the vortex pair can be uniquely unwrapped, provided the de Broglie wavelength is well resolved. However, the $\pi$ phase jump along this line poses numerical difficulties for fluid-based schemes (see also \fref{fig:vortex_phase}). See \citet{Hui2021a} for a variety of vortex configurations.

\subsection{Tests with self-gravity}
\label{subsec:tests_with_gravity}

\subsubsection{Jeans instability}
\label{subsec:tests_jeans}

In analogy to the Jeans instability in hydrodynamics, quantum pressure in FDM gives rise to a characteristic Jeans scale. Perturbations on scales larger than this threshold grow under gravity, whereas those on smaller scales are stabilized by quantum pressure. To examine the linear evolution of a perturbed wave function in a homogeneous background, one can insert $\psi=(1+\dR) + i\dI$ into Eqs. (\ref{eq:schroedinger}--\ref{eq:poisson}) and assume $\dR \ll 1$ and $\dI \ll 1$, with the background density normalized to unity. By retaining the leading-order terms and applying the conventional Jeans swindle, the spatial Fourier components of the wave function satisfy
\begin{align}
  \frac{{\rm d}^2 \dR_k}{{\rm d} t^2} &= \frac{\hbar^2}{4m^2}\left( \kJphy^4 - k^4 \right) \dR_k,
  \label{eq:jeans_phy_real} \\
  \dI_k &= \frac{2m}{\hbar k^2} \frac{{\rm d} \dR_k}{{\rm d} t},
  \label{eq:jeans_phy_imag}
\end{align}
where $k$ is the wavenumber and
\begin{equation}
  \kJphy = \left( \frac{16\pi Gm^2}{\hbar^2} \right)^{1/4}
  \label{eq:jeans_phy_wavenumber}
\end{equation}
is the Jeans wavenumber in physical coordinates (note that $\kJ$ refers to the \emph{comoving} Jeans wavenumber throughout this article). The corresponding solutions are
\begin{alignat}{2}
  \dR(x,t) &= \dR_0 \cos(kx + \theta_0) {\rm e}^{\pm \omega_1 t}, \quad & k<\kJphy,
  \label{eq:jeans_phy_real_sol_unstable} \\
  \dR(x,t) &= \dR_0 \cos(kx \pm \omega_2 t + \theta_0), \quad & k>\kJphy,
  \label{eq:jeans_phy_real_sol_stable}
\end{alignat}
where $\omega_1=(\hbar/2m)(\kJphy^4-k^4)^{1/2}$, $\omega_2=(\hbar/2m)(k^4-\kJphy^4)^{1/2}$, and $\dR_0$ and $\theta_0$ are arbitrary constants.

\begin{figure*}[th!]
\centering
\includegraphics[width=0.6\textwidth]{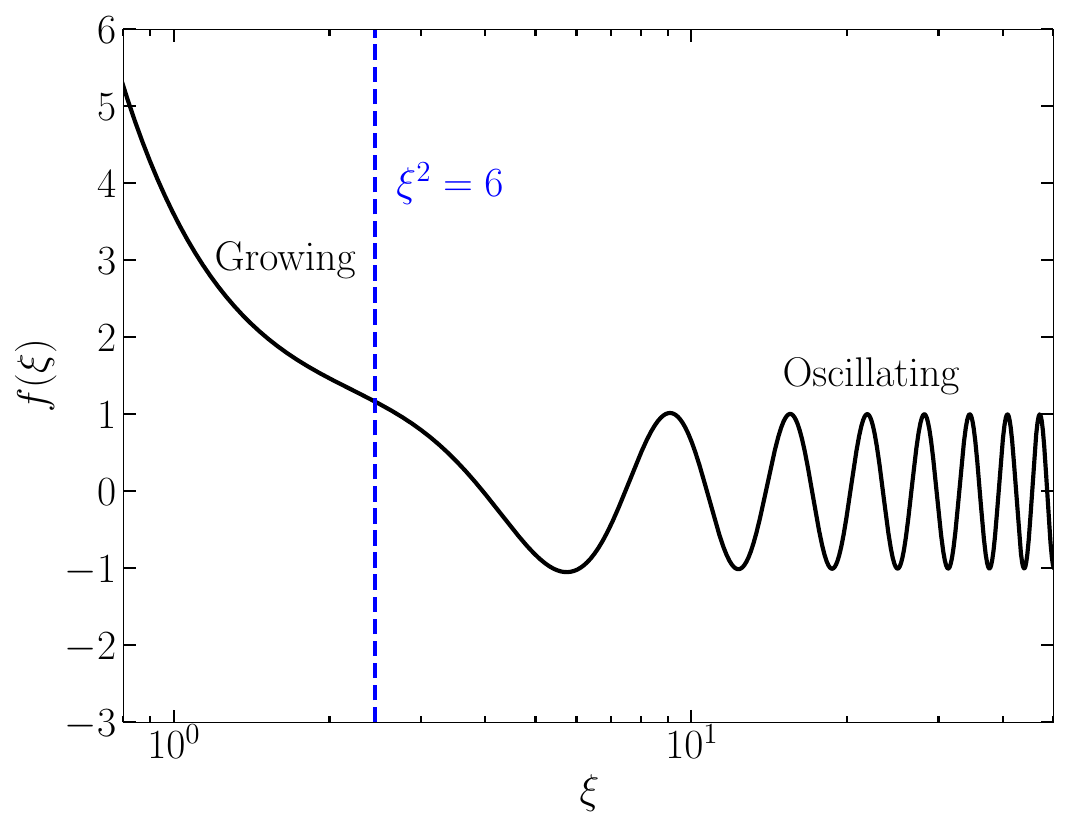}
\caption{
Analytical solution $f(\xi)$ (see \erefp{eq:jeans_com_real_sol_grow}) to the comoving Jeans instability equation (\erefp{eq:jeans_com_real}), where $\xi \propto k^2 a^{-1/2}$. The plot illustrates the transition from growth to oscillation at the comoving Jeans wavenumber $\kJ$, which corresponds to $\xi^2 = 6$ (vertical line; see \erefp{eq:jeans_com_wavenumber})
}
\label{fig:jeans}
\end{figure*}

Similarly, in comoving coordinates, one can insert $\tilde{\psi}=(1+\dRcom) + i\dIcom$ into Eqs. (\ref{eq:schroedinger_com}--\ref{eq:poisson_com}) to obtain the following governing equations for the spatial Fourier components of the comoving wave function \citep{Woo2009}:
\begin{align}
  \xi^2\frac{{\rm d}^2 \dRcom_k}{{\rm d} \xi^2} &= (6-\xi^2) \dRcom_k,
  \label{eq:jeans_com_real} \\
  \dIcom_k &= -\frac{{\rm d} \dRcom_k}{{\rm d} \xi},
  \label{eq:jeans_com_imag}
\end{align}
where
\begin{equation}
  \xi = \frac{\hbar}{mH_0}\frac{k^2}{a^{1/2}}.
  \label{eq:jeans_com_xi}
\end{equation}
$H_0$ denotes the present-day Hubble parameter. Here we assume a matter-dominated universe with matter density parameter $\Omega_m=1$ and normalize the background density $\rhob$ to unity. \eref{eq:jeans_com_real} admits two independent solutions:
\begin{align}
  f(\xi) &= \frac{3\cos\xi + 3\xi\sin\xi - \xi^2\cos\xi}{\xi^2},
  \label{eq:jeans_com_real_sol_grow} \\
  g(\xi) &= \frac{3\sin\xi - 3\xi\cos\xi - \xi^2\sin\xi}{\xi^2}.
  \label{eq:jeans_com_real_sol_decay}
\end{align}
Furthermore, a characteristic length scale arises at $\xi^2=6$, which defines the comoving Jeans wavenumber \citep{Hu2000}:
\begin{equation}
  \kJ = \left( \frac{6aH_0^2m^2}{\hbar^2} \right)^{1/4} = a^{1/4}\kJphy.
  \label{eq:jeans_com_wavenumber}
\end{equation}
For $k \ll \kJ$, $f(\xi) \propto a$ and $g(\xi) \propto a^{-3/2}$, corresponding to the growing and decaying modes, respectively. For $k \gg \kJ$, both $f(\xi)$ and $g(\xi)$ exhibit oscillatory behavior. \fref{fig:jeans} illustrates $f(\xi)$ and the transition at $\xi^2 = 6$.

The Jeans instability setup is useful for quantifying the error convergence rates of both fluid-based and wave-based schemes involving self-gravity. The corresponding solutions for fluid variables can be derived straightforwardly from Eqs. (\ref{eq:mass_dens}--\ref{eq:engy_dens}) \citep[see also][]{Woo2009}. There is, however, one caveat. The solutions given in Eqs. (\ref{eq:jeans_com_real_sol_grow}--\ref{eq:jeans_com_real_sol_decay}) rely on the assumption of $\dIcom^2 \ll 2\dRcom$ in the source term of the Poisson equation. This assumption can break down rapidly in the growing-mode solution with $\xi \ll 1$, since $\| \dIcom/\dRcom \| = 2/\xi \gg 1$ and this ratio increases as $a^{1/2}$.

\subsubsection{Solitons}
\label{subsec:tests_soliton}

\begin{figure*}[th!]
\centering
\includegraphics[width=\textwidth]{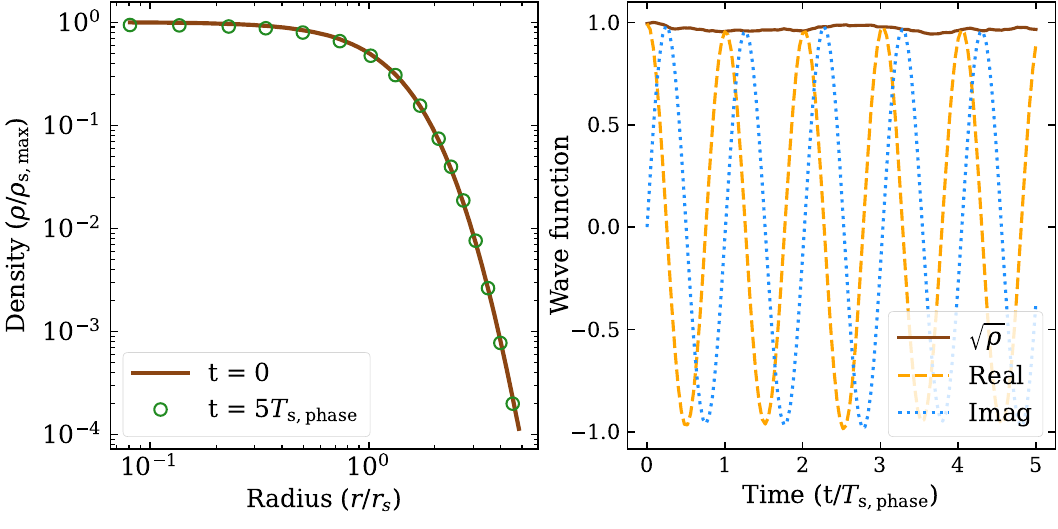}
\caption{
Test of an isolated soliton. (Left) Agreement between the initial and final density profiles after $5\,T_{\rm s, phase}$. (Right) Evolution of the wave function at the soliton center. The central density remains approximately constant, while the real and imaginary parts oscillate with a period consistent with \eref{eq:soliton_period_phase}
}
\label{fig:test_soliton}
\end{figure*}

Solitons are spherically symmetric, non-dispersive, ground-state solutions of the self-gravitating Schr\"odinger--Poisson system (see \sref{subsec:soliton}). Their density profiles feature a flat core supported by quantum pressure, followed by a steep monotonic decrease (see \fref{fig:test_soliton}). Furthermore, these solutions obey the scaling symmetry described in \eref{eq:soliton_scale_transf}.

The soliton solutions can be obtained by inserting the stationary ansatz $\psi(r,t)=e^{-i\omega t}\Psi(r)$ into Eqs. (\ref{eq:schroedinger}--\ref{eq:poisson}), where $\omega$ is the angular frequency and $\Psi$ is real. This form implies a coherent phase and zero bulk velocity. The substitution yields a coupled system of second-order ordinary differential equations, which can be solved numerically with appropriate boundary conditions by seeking the lowest-energy, node-free solution \citep{Guzman2004, Marsh2015a}. Although no exact analytical solution exists, numerical solutions are well approximated by \eref{eq:soliton_profile} within $\sim 3\,\rs$, enclosing approximately $95\%$ of the total soliton mass \citep{Schive2014a}.

This setup is particularly suited for validating numerical accuracy in the fully nonlinear regime. As an illustration, \fref{fig:test_soliton} shows simulation results obtained using the global Fourier method over a domain of $10\,\rs$ with spatial resolution $\Dh \approx 0.08\,\rs$. The simulation is evolved for $5\,T_{\rm s, phase}$, where
\begin{equation}
  T_{\rm s, phase} \approx 38.2 \left( \frac{\rho_{\rm s, max}}{\Msun\pc^{-3}} \right)^{-1/2} \Myr
  \label{eq:soliton_period_phase}
\end{equation}
is the oscillation period of the soliton phase, with $\rho_{\rm s, max}$ denoting the maximum soliton density \citep{Guzman2004, Veltmaat2018, Chiang2021}. The initial and final density profiles match closely. The central density exhibits small-amplitude oscillations, primarily due to the limited accuracy of the fitting function \eref{eq:soliton_profile} beyond $3\,\rs$. The phase of the wave function oscillates with a period consistent with \eref{eq:soliton_period_phase}.

In addition to testing a stationary soliton solution, one can also simulate a perturbed configuration. In such cases, the central density undergoes periodic oscillations with a characteristic timescale given by \citep{Guzman2004, Veltmaat2018, Chiang2021}
\begin{equation}
  T_{\rm s, density} \approx 92.1 \left( \frac{\rho_{\rm s, max}}{\Msun\pc^{-3}} \right)^{-1/2} \Myr
                     \approx 2.4\,T_{\rm s, phase}.
  \label{eq:soliton_period_dens}
\end{equation}
A perturbed soliton with an oscillation amplitude of order unity is commonly observed in soliton--halo systems forming in cosmological simulations. This behavior can be interpreted as the superposition of the ground-state soliton and excited states \citep{Li2021}.

The stable soliton configuration offers a convenient simulation testbed for exploring various properties of FDM. Example applications include studies of soliton--halo systems through soliton mergers \citep{Schive2014b, Schwabe2016, Mocz2017, Li2021, Chan2024, Stallovits2025, Blum2025}, tidal stripping of solitons \citep{Du2018}, dynamical interactions between solitons and stars \citep{Chan2018, Chiang2021}, dynamical friction of massive objects within solitons \citep{Wang2022, Boey2024, Boey2025}, modeling the central molecular zone of the Milky Way \citep{Li2020}, and investigating the enhanced growth of supermassive black holes at high redshifts \citep{Chiu2025}.

\subsubsection{Isolated halos}
\label{subsec:tests_halo}

An isolated halo serves as a stringent benchmark for testing FDM simulations. A robust FDM algorithm, whether wave- or fluid-based, must meet the following criteria. The halo density profile within the virial radius should be stable on a dynamical timescale. The total mass, momentum, and energy must be conserved with errors below the few-percent level. The kinetic and thermal energies (Eqs. \ref{eq:bulk_vel}--\ref{eq:engy_dens}) are expected to exhibit energy equipartition, and their sum should follow the virial condition. The density granules need to be well resolved throughout the entire halo.

The central soliton core must also be well resolved and remain stable. In a typical system, the soliton peak density exceeds its ambient halo density by more than an order of magnitude \citep{Schive2014a, Schive2014b}, with a sharp soliton--halo transition at $r \approx 3.3\text{--}3.5\,\rs$ \citep{Mocz2017, Chiang2021}. Within this radius, the density profile should closely follow \eref{eq:soliton_profile} and be supported by quantum pressure, with negligible turbulent motion. The soliton peak density is expected to oscillate with an amplitude of order unity and a characteristic period given by \eref{eq:soliton_period_dens}. In addition, the soliton undergoes a confined random walk with respect to the halo center of mass, with a characteristic displacement and timescale on the order of $\rs$ and $T_{\rm s, phase}$, respectively \citep{Schive2020, Li2021, Chowdhury2021}. Over longer timescales comparable to the condensation time, the soliton mass may exhibit secular growth \citep{Levkov2018, Eggemeier2019, Chen2021, Chan2024}.

The isolated halo test provides a valuable means of calibrating the resolution requirements for an FDM algorithm in a fully nonlinear, self-gravitating system. Specifically, it can be used to determine the optimal spatial and temporal resolution needed to resolve the characteristic length and time scales of the ubiquitous density granules, as inferred from Eqs. (\ref{eq:granule_size}--\ref{eq:granule_time}). In principle, these resolution criteria are broadly applicable to other FDM simulations. An additional complication arises for Lagrangian approaches due to their inherently fixed mass resolution---they may lack sufficient spatial resolution to resolve density granules and quantum pressure in the low-density halo outskirts (see \sref{subsec:challenges_wavelength}). Furthermore, the phase discontinuity and the divergence of velocity and quantum potential near vortices could present significant numerical challenges for fluid-based schemes (see \sref{subsec:challenges_vortices}). It also remains an open question whether fluid schemes can accurately capture the soliton core in a soliton--halo system, characterized by a high-density plateau, a sharp soliton--halo transition, and negligible velocity dispersion \citep{Nori2021, Nori2023, Chan2025}.

\begin{figure*}[th!]
\centering
\includegraphics[width=\textwidth]{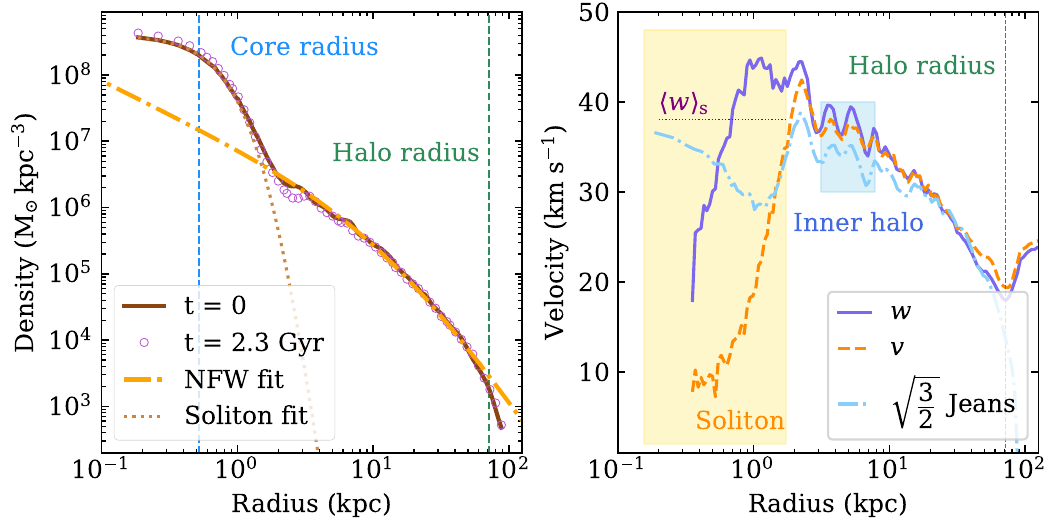}
\caption{
Test of an isolated halo with $\Mh=1.9\times10^{10} \Msun$ and $\mFDM=0.8$. (Left) Density profiles at the beginning and after evolving for approximately one free-fall time, demonstrating halo stability. The central core aligns well with the soliton solution, while the outer region follows the NFW model. (Right) Velocity profiles. Outside the soliton, the thermal velocity $w$ and bulk velocity $v$ coincide due to energy equipartition, and both follow the profile derived from the isotropic spherical Jeans equation (dash-dotted line). Inside the soliton (yellow shaded region), the thermal velocity dominates, and its average value $\ave{w}_s$ matches that of the inner halo (blue shaded region), indicating thermal equilibrium. See also \fref{fig:test_halo_slice}
}
\label{fig:test_halo_profile}
\end{figure*}

As a demonstration, \fref{fig:test_halo_profile} shows the density and velocity profiles of an FDM halo extracted from a cosmological simulation with $\mFDM=0.8$ at $z=0$. We extract a cubic region of width $186 \kpc$, centered on a halo with a virial radius $\rh=72 \kpc$ and a mass $\Mh=1.9\times10^{10} \Msun$ (see \nameref{sec:data} for the link to download the halo data). The central soliton has a radius $\rs=0.52 \kpc$ and a mass $\Ms=1.5\times10^8 \Msun$, with a soliton--halo transition approximately at $4\,\rs$. The halo density profile at $r \gtrsim 8\,\rs$ is well fitted by the NFW model with a concentration parameter of $c=5.4$. To validate the system's stability, we evolve it for another $2.3 \Gyr$ (roughly one free-fall time), using the global Fourier method employed in the \textcode{GAMER} code on a $N=1024^3$ grid without AMR, corresponding to a spatial resolution of $0.18 \kpc$. We apply isolated boundary conditions for the Poisson solver and periodic boundary conditions for the wave function.

The left panel demonstrates that the density profile outside the soliton remains stable throughout the evolution. The central core closely follows the soliton solution \eref{eq:soliton_profile}, with minor differences between the initial and final profiles attributable to stochastic soliton oscillations. The right panel shows that the thermal and bulk velocities in the inner halo are closely aligned, indicating energy equipartition. Both velocities match well the velocity dispersion predicted by the isotropic spherical Jeans equation \citep{Chowdhury2021}. Furthermore, within the soliton, the bulk velocity is negligible, while the average thermal velocity $\ave{w}_{\rm s}$ approximately matches the inner-halo thermal velocity, suggesting thermal equilibrium between the soliton and its ambient halo. See \citet{Liao2025} for further discussion.

\begin{figure*}[th!]
\centering
\includegraphics[width=\textwidth]{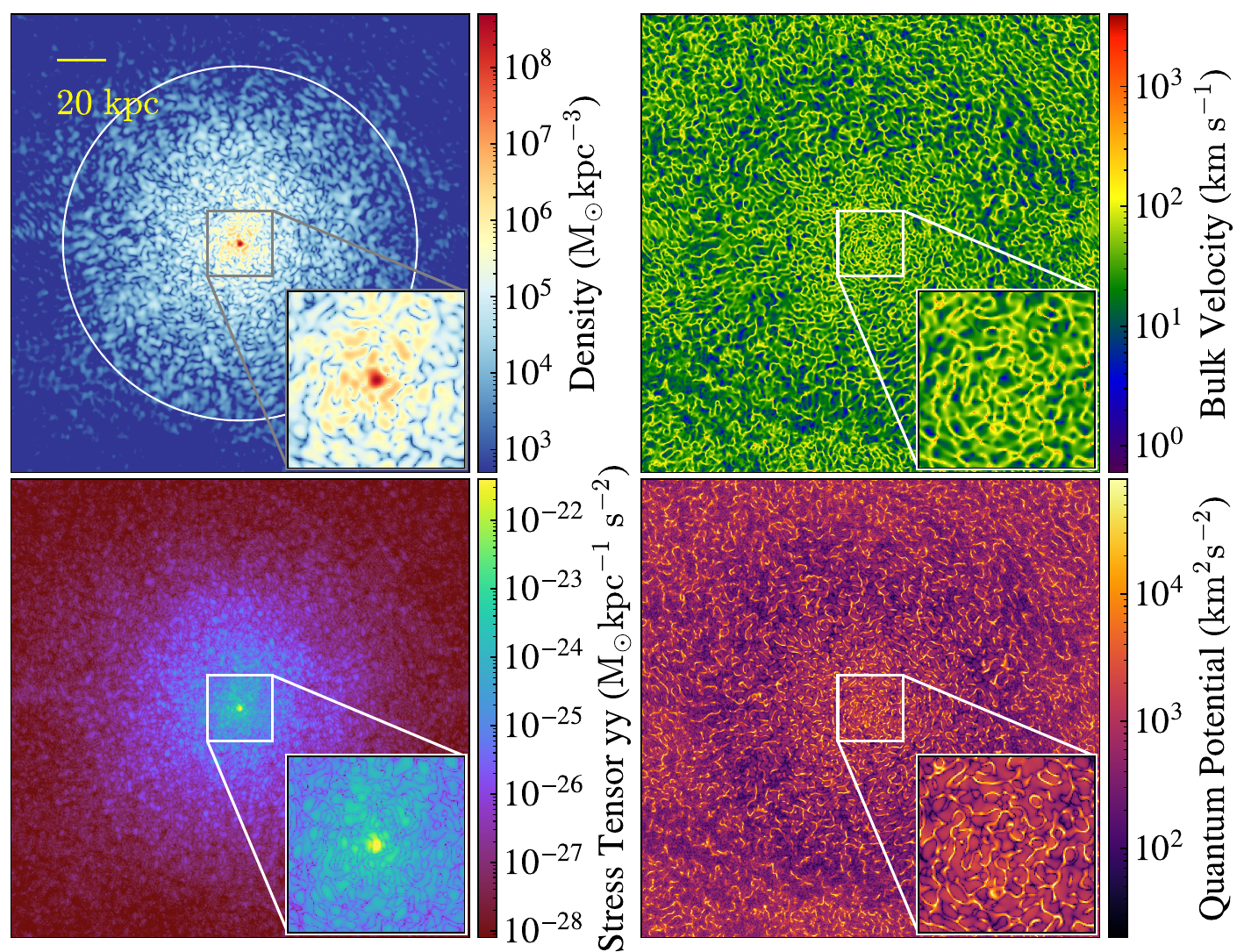}
\caption{
Test of the same isolated halo shown in \fref{fig:test_halo_profile}. The panels show slices of mass density $\rho$ (upper left), bulk velocity $v$ (upper right), stress tensor $\Sigma_{yy}$ (lower left), and quantum potential $Q$ (lower right) through the halo center. $\rho$ and $\Sigma_{yy}$ peak at the central soliton, whereas $v$ and $Q$ display filamentary structures and reach extreme values at isolated points, corresponding to vortices in density voids. The solid circle indicates the halo virial radius
}
\label{fig:test_halo_slice}
\end{figure*}

\fref{fig:test_halo_slice} shows the mass density $\rho$, bulk velocity magnitude $v$, quantum stress tensor $\Sigma_{yy}$ (see \erefp{eq:quantum_pressure}), and quantum potential $Q$ (see \erefp{eq:quantum_potential}) on a slice through the soliton center. Both $\rho$ and $\Sigma_{yy}$ exhibit relatively smooth distributions and peak at the central soliton. By contrast, $v$ and $Q$ show filamentary structures that trace low-density voids caused by destructive interference. Moreover, there are isolated points with extremely high values of $v$ and $Q$, corresponding to locations where vortex lines intersect the plotting plane (see \sref{subsec:challenges_vortices}).

In addition to being extracted from cosmological simulations, isolated FDM halos can also be constructed from scratch via the gravitational collapse of a bound system \citep[e.g.,][]{Li2021, Hui2021a, Blum2025} or from a prescribed distribution function (see \sref{subsec:methods_eigen}). Furthermore, beyond their use in calibrating FDM algorithms, isolated halos serve as versatile simulation testbeds for investigating several distinctive FDM phenomena, such as the formation and dynamics of quantized vortices \citep{Hui2021a}, the stochastic motion of nuclear objects \citep{Schive2020, Chowdhury2021, Li2021}, and dynamical heating in dwarf galaxies \citep{Dalal2022, Chowdhury2023, Teodori2026}, galactic disks \citep{Yang2024}, and stellar streams \citep{Dalal2021}.

\subsubsection{Cosmological simulations}
\label{subsec:tests_cosmo}

Cosmological simulations arguably provide the most comprehensive test for FDM algorithms. For wave-based schemes, sufficient resolution to resolve the de Broglie wavelength must be maintained throughout the entire computational domain. Simulations lacking adequate resolution in large-scale, low-density regions outside halos may misestimate the flow velocity, thereby biasing halo formation and merger histories. Within halos, insufficient resolution may lead to errors in the quantum pressure and turbulence, resulting in incorrect density profiles, soliton properties, and dynamical heating effects from density granules. See \sref{subsec:challenges_wavelength} for related discussion. In addition, mass conservation error provides a useful diagnostic of numerical accuracy for non-conservative numerical schemes, such as wave-based finite-difference and local pseudo-spectral methods. For AMR, these resolution requirements help calibrate grid refinement criteria, which are nontrivial and depend on the adopted evolution schemes. Furthermore, numerical artifacts at coarse--fine interfaces must be carefully assessed and minimized.

Fluid-based schemes can accurately capture large-scale structure with reduced resolution requirements, as they do not need to resolve the de Broglie wavelength. However, as discussed in Sections \ref{subsec:challenges_wavelength} and \ref{subsec:tests_halo}, it remains debated whether these schemes can reliably capture substructures within FDM halos, such as density granules, vortices, and solitons. For hybrid approaches, it is essential to monitor and minimize numerical artifacts at fluid--wave interfaces, ensure consistency with high-resolution wave-only simulations, and demonstrate superior performance compared to those simulations.

Validation of numerical convergence with increasing resolution can be challenging, and sometimes prohibitively expensive, for FDM cosmological simulations, especially for wave schemes with higher $\mFDM$ or larger simulation volumes. This difficulty arises primarily from the need to (i) use smaller time steps ($\Dt \propto 1/\mFDM \vdisp^2$) and (ii) apply higher spatial resolution ($\Dh \propto 1/\mFDM\vdisp$) across a larger fraction of the simulation domain than in CDM simulations. An added complication is the potential formation of spurious halos (see \sref{subsec:challenges_spurious_halos}). A practical and efficient alternative for assessing numerical convergence is to perform collisionless $N$-body simulations using the same initial conditions \citep{Liao2025}. First, the halo masses and the density profiles outside the central solitons in the genuine FDM and $N$-body simulations should agree closely. Note that a good fit to an NFW profile beyond the soliton does not guarantee numerical accuracy, as an incorrect profile may still match an NFW model but with erroneous halo mass and concentration. Second, the halo positions should be consistent across the two types of simulations, since the spatial distribution of spurious halos is primarily driven by numerical noise (see \fref{fig:spurious_halo}).

To obtain the initial conditions for FDM cosmological simulations, one can first use the Boltzmann code \textcode{axionCAMB} to compute the FDM initial power spectrum. Subsequently, one can use an initial condition tool such as \textcode{MUSIC} \citep{Hahn2011}, \textcode{N-GenIC} \citep{Springel2015}, or \textcode{MPgrafic} \citep{Prunet2008} to construct the three-dimensional comoving mass density $\tilde{\rho}$ and comoving peculiar velocity $\bm{\tilde{v}} = a\bm{v_{\rm pec}}$. For wave-based schemes and fluid-based schemes that evolve the phase field, the comoving wave function is given by $\tilde{\psi} = \tilde{\rho}^{1/2} e^{i \tilde{S}}$, where $\bm{\tilde{v}} = (\hbar/m)\bm{\tilde{\nabla}}\tilde{S}$ (see \sref{subsec:intro_eq}). Accordingly, the initial wave function can be constructed by solving
\begin{align}
  \| \tilde{\psi} \| &= \tilde{\rho}^{1/2},
  \label{eq:cosmo_ic_amp} \\
  \tilde{\nabla}^2 \tilde{S} &= \frac{m}{\hbar} \bm{\tilde{\nabla}} \cdot \bm{\tilde{v}}.
  \label{eq:cosmo_ic_phase}
\end{align}

\begin{figure*}[th!]
\centering
\includegraphics[width=\textwidth]{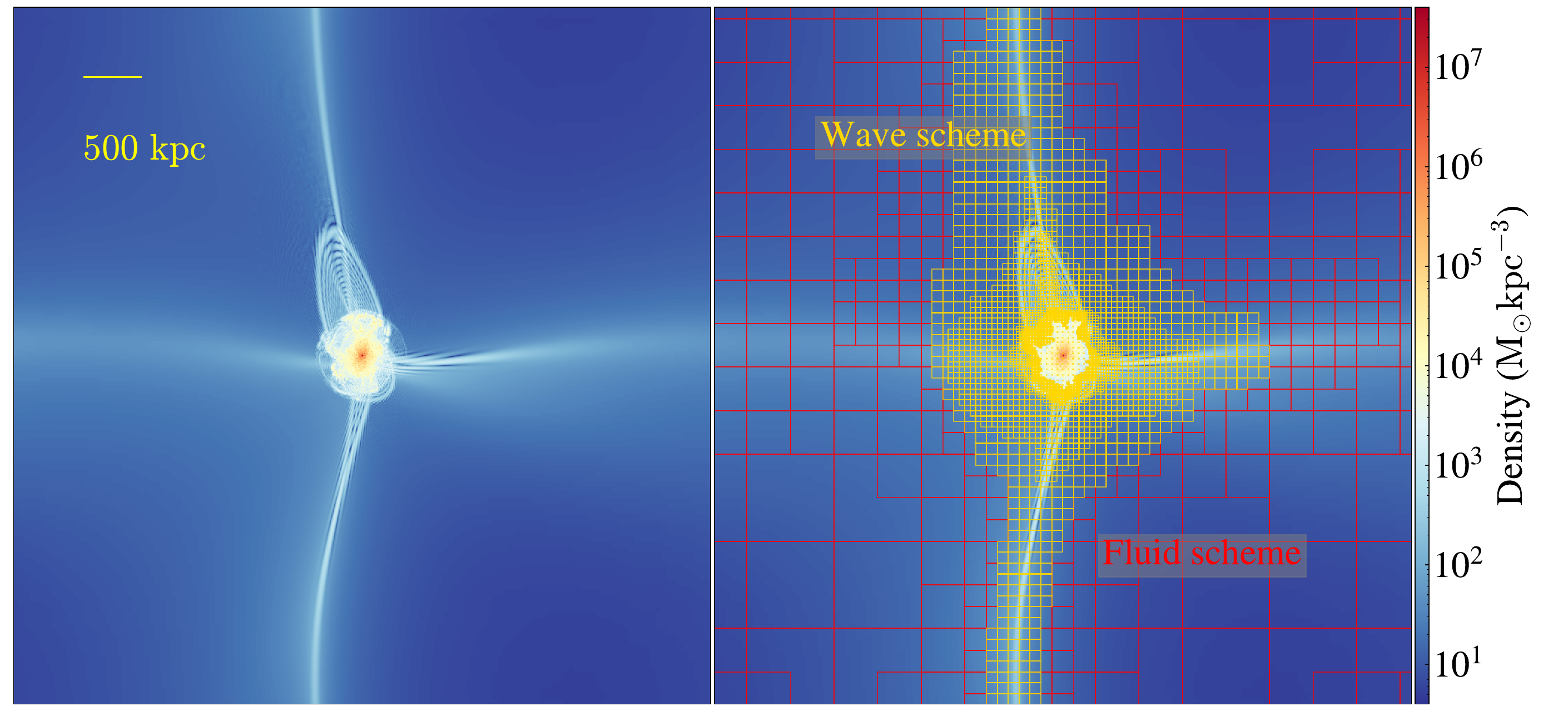}
\caption{
Cosmological simulation in a comoving box of size $L = 4\Mpch$ with $\mFDM = 0.1$ at $z=0$. (Left) Projected density map centered on a halo of mass $\Mh = 5.9\times10^{11}\Msunh$. (Right) AMR grid map showing the regions that employ the fluid and wave schemes, respectively, in the hybrid algorithm of \citet{Kunkel2025}
}
\label{fig:test_cosmo_slice}
\end{figure*}

\begin{figure*}[th!]
\centering
\includegraphics[width=0.8\textwidth]{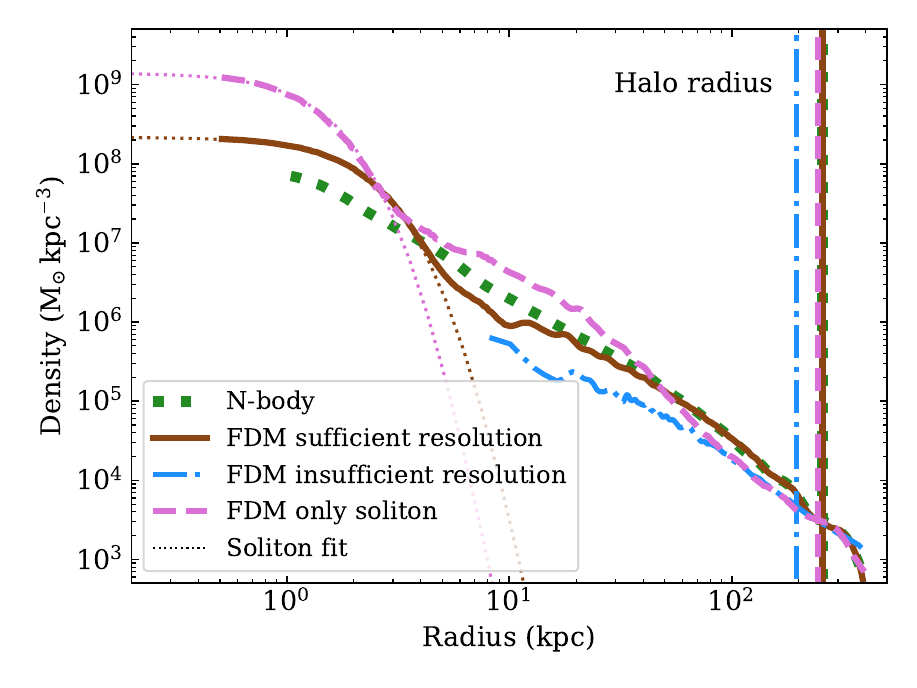}
\caption{
Density profiles of the central halo in \fref{fig:test_cosmo_slice} at $z=0$ obtained with different evolution schemes from the same initial condition. For the FDM simulation using a hybrid scheme with sufficient resolution (solid line), the central core is well fitted by the soliton solution (thin dotted lines), and the outer profile matches the collisionless $N$-body simulation (thick dotted line). When the same hybrid scheme is applied but with adequate wave-scheme resolution only inside the soliton (dashed line), both the central soliton and the outer profile become overly concentrated, although the halo mass remains accurate, as indicated by the corresponding halo virial radius (vertical lines). In contrast, using the global Fourier method with insufficient resolution to resolve the de Broglie wavelength severely underestimates the halo mass (dash-dotted line)
}
\label{fig:test_cosmo_profile}
\end{figure*}

As an illustration, Figs. \ref{fig:test_cosmo_slice} and \ref{fig:test_cosmo_profile} show results from FDM cosmological simulations with $\mFDM = 0.1$ at $z=0$ in a comoving box of size $L = 4\Mpch$. See \nameref{sec:data} for the link to download the initial condition files. \fref{fig:test_cosmo_slice} shows the projected mass density and AMR grid distribution from a simulation using the \textcode{GAMER} code with a hybrid scheme \citep{Kunkel2025}. This simulation employs a $64^3$ root grid and 8 additional refinement levels, achieving a maximum resolution of $0.24\kpch$. The fluid scheme that solves the Hamilton--Jacobi--Madelung equations is applied at levels $0\text{--}3$, which cover smooth, low-density regions and occupy the majority of the simulation volume. In comparison, the wave scheme based on the FC--Gram algorithm is applied at levels $4\text{--}8$, which target regions with strong interference, such as filaments and halos, and dominate the computational cost.

\fref{fig:test_cosmo_profile} compares the density profiles of the central halo in \fref{fig:test_cosmo_slice} at $z=0$ obtained with different evolution schemes. This halo has a mass $\Mh = 5.9\times10^{11}\Msunh$ and a virial radius $\rh = 166\kpch$. The central soliton has a mass $\Ms = 2.2\times10^9\Msunh$ and a radius $\rs = 1.1\kpch$. The free-fall velocity outside the halo is approximately $110\kms$, corresponding to a de Broglie wavelength of $\lambda_{\rm dB, ff} \sim 7.6\kpch$. For the hybrid scheme with fiducial resolution, the central profile aligns well with the soliton solution (\erefp{eq:soliton_profile}), and the outer profile closely matches that of a collisionless $N$-body simulation using the \textcode{GADGET-2} code with the same initial condition. In contrast, the global Fourier method, performed with a $1024^3$ grid and $3.9\kpch$ spatial resolution, underestimates the halo mass because the resolution is insufficient to properly resolve $\lambda_{\rm dB, ff}$. When the same hybrid scheme is applied with adequate resolution confined to the soliton but inadequate resolution elsewhere in the halo, quantum pressure and turbulence are underestimated, leading to unphysical halo contraction. The deepened gravitational potential raises the halo temperature and, in turn, increases the soliton energy and mass. Nevertheless, note that (i) the resulting central profile still fits a soliton solution, and (ii) the halo mass remains in good agreement with the fiducial value. The latter is attributed to the fluid scheme employed outside the halo. These observations suggest that agreement in halo mass and a good match to the theoretical soliton profile are necessary but not sufficient to establish numerical accuracy \citep{Liao2025}.

Due to the extremely high computational cost, most FDM cosmological simulations to date are limited to probing relatively small FDM particle masses ($\mFDM \sim 0.1\text{--}2$)---either in small simulation boxes ($L \sim 1\text{--}2\Mpch$) evolved to $z \sim 0\text{--}1$, or in somewhat larger boxes ($L \sim 5\text{--}20\Mpch$) but only to higher redshifts ($z \gtrsim 3$). Representative examples include simulations using the global Fourier method, applied to either pure FDM scenarios \citep{Woo2009, Li2019, May2021, May2022, Chan2022} or a mixed CDM--FDM model \citep{Lague2024}, and the wave-based finite-difference scheme with AMR \citep{Schive2014a, Schive2014b}. Simulations employing hybrid schemes and zoom-in techniques enable resolving more massive halos, simulating larger volumes, or reaching lower redshifts \citep{Veltmaat2018, Schwabe2022, Kunkel2025, Chan2025, Chiu2025}. In comparison, $N$-body \citep{Schive2016, Sarkar2016, Irsic2017, Corasaniti2017, Armengaud2017, Kobayashi2017, Ni2019, Leong2019, Nadler2024} and SPH \citep{Zhang2018b, Nori2019, Nori2021, Nori2023} simulations can explore much higher $\mFDM$ and larger volumes at lower redshifts, albeit each with its own limitations. A few pioneering simulations have incorporated baryonic physics \citep{Mocz2019, Mocz2020, Veltmaat2020}, though they are currently restricted to $\mFDM=2.5$, $L\sim2\Mpch$, and $z\sim4\text{--}5$.

\section{Conclusions}
\label{sec:conclusions}

In this review, we introduced a variety of FDM algorithms (\sref{sec:methods}), described the associated numerical challenges (\sref{sec:challenges}), and presented a representative set of numerical tests (\sref{sec:test}). Wave-based methods most accurately capture the fine-grained wave phenomena in FDM halos, such as density granulation, vortices, and soliton cores (\sref{subsec:fdm_features}). However, they are computationally expensive due to the need to resolve the de Broglie wavelength and its rapid oscillations. Moreover, they do not, in general, guarantee conservation of momentum and energy.

Fluid-based methods are significantly more efficient for modeling smooth, high-velocity flows. They ensure manifest conservation and can readily adopt a Lagrangian formulation that preserves Galilean invariance. However, they struggle to resolve wave features in interference-dominated, multi-stream regions. Hybrid schemes, coupled with AMR, apply fluid methods in smooth, single-stream regions and switch to wave methods in complex, multi-stream regions, thereby enabling simulations of larger volumes and higher FDM particle mass. Nevertheless, simulating halos more massive than $\sim 10^{12}\Msun$ to $z \sim 0$ with $\mFDM \gg 1$ likely remains computationally infeasible in the near future.

Several alternative approaches are also worth noting. Eigenmode methods offer efficient tools for constructing FDM halos or simulating restricted halo regions, provided that gravitational backreaction from non-dark-matter components can be neglected. Collisionless $N$-body methods, while neglecting quantum pressure, remain valuable for probing large-scale structure and assessing the numerical convergence of genuine FDM cosmological simulations. Finally, machine learning, though not addressed in this review, offers an interesting direction for accelerating or augmenting FDM simulations \citep[e.g.,][]{Mishra2025}.

The discussions in this review have mostly focused on a single FDM species without self-interaction. However, the algorithms described here can be easily extended to explore more general scenarios---for example, FDM with self-interaction \citep[e.g.,][]{Mocz2023, Jain2023, Painter2024, Glennon2024, Stallovits2025}, vector or multi-component FDM \citep[e.g.,][]{Amin2022, Huang2023, Gosenca2023, Luu2024}, and mixed CDM--FDM models \citep[e.g.,][]{Schwabe2020, Lague2024}.

To facilitate comparison among different FDM codes, we provide links to the initial condition files for the isolated-halo and cosmological simulations in \nameref{sec:data}.

\addcontentsline{toc}{section}{Data availability}
\section*{Data availability}
\label{sec:data}

The initial condition files of the numerical tests presented in Sections \ref{subsec:tests_halo} (isolated halos) and \ref{subsec:tests_cosmo} (cosmological simulations) can be downloaded from the link: \url{https://drive.google.com/drive/folders/1Xjgi6AYtl8aNhqLplPhnt_FVSooyBQeA?usp=sharing}.

\vspace{0.5cm}
\backmatter
\bmhead{Acknowledgments}

I would like to thank Pin-Yu Liao for polishing the figures and Hsinhao Huang, Alexander Kunkel, and Guan-Ming Su for their helpful comments. We use \textcode{yt} \citep{yt} for data visualization and analysis. This research is partially supported by the National Science and Technology Council (NSTC) of Taiwan under Grant No. NSTC 111-2628-M-002-005-MY4 and the NTU Academic Research--Career Development Project under Grant No. NTU-CDP-113L7729.

\section*{Declarations}
\bmhead{Conflict of interest}
The author declares no conflict of interest.



\newcommand {\apj}      {Astrophys. J. }
\newcommand {\apjl}     {Astrophys. J. Lett. }
\newcommand {\apjs}     {Astrophys. J. Suppl. }
\newcommand {\mnras}    {Mon. Not. R. Astron. Soc. }
\newcommand {\mnrasl}   {Mon. Not. R. Astron. Soc. Lett. }
\newcommand {\aap}      {Astron. Astrophys. }
\newcommand {\jcap}     {J. Cosmol. Astropart. Phys. }
\newcommand {\na}       {New Astron. }
\newcommand {\nar}      {New Astron. Rev. }
\newcommand {\araa}     {Annu. Rev. Astron. Astrophys. }
\newcommand {\physrep}  {Phys. Rep. }
\newcommand {\prl}      {Phys. Rev. Lett. }
\newcommand {\prd}      {Phys. Rev. D }
\newcommand {\pre}      {Phys. Rev. E }
\newcommand {\jhep}     {J. High Energy Phys. }
\newcommand {\jcp}      {J. Comput. Phys. }
\newcommand {\naa}      {Nat. Astron. }
\newcommand {\nap}      {Nat. Phys. }
\newcommand {\lrca}     {Living Rev. Comput. Astrophys. }
\newcommand {\ppnp}     {Prog. Part. Nucl. Phys. }
\newcommand {\fass}     {Front. Astron. Space Sci. }
\newcommand {\pr}       {Phys. Rev. }
\newcommand {\epjp}     {Eur. Phys. J. Plus }
\newcommand {\epjc}     {Eur. Phys. J. C }

\newpage
\phantomsection
\addcontentsline{toc}{section}{References}
\bibliography{references.bib}


\end{document}